\documentclass[12pt]{article}
\usepackage{amsmath, amssymb, graphics}
\usepackage{graphicx}
\usepackage{bbm}
\usepackage{color}
\usepackage{subfigure}
\usepackage{float}
\usepackage{appendix}

\usepackage{graphics, psfrag}
\newcommand{\mathsym}[1]{{}}
\def\id{\protect{{1 \kern-.28em {\rm l}}}}

\def\be{\begin{eqnarray}}
\def\ee{\end{eqnarray}}

\makeatletter
\renewcommand\section{\@startsection {section}{1}{\z@}%
                                   {-3.5ex \@plus -1ex \@minus -.2ex}%
                                   {2.3ex \@plus.2ex}%
                                   {\normalfont\large\bfseries}}
\renewcommand\subsection{\@startsection{subsection}{2}{\z@}%
                                   {-3.25ex\@plus -1ex \@minus -.2ex}%
                                   {1.5ex \@plus .2ex}%
                                   {\normalfont\normalsize\bfseries}}

\makeatother

\def \del{\partial}

\def\s{\sigma}

\def\ov{\over}

\def \la {\label}

\def \l {\lambda}

\def \sql {{\sqrt \l}}
\def \adss {$AdS_5 \times S^5~$ }

\def \ov {\over}

\def\th{\theta}

\def\adss{AdS$_5\times$S$^5$}

\def\NeqFour{{\cal N}=4}

\def\rx{{\rm x}}
\def\ry{{\rm y}}

\def\V{{\rm V}}
\def\D{\Delta}
\def\del{\partial}
\def\bd{{\bar\partial}}
\def\rY{{\rm Y}}

\def\ha{{\textstyle{\frac{1}{2}}}}

\numberwithin{equation}{section}

\begin{document}


\overfullrule=0pt
\parskip=2pt
\parindent=12pt
\headheight=0in \headsep=0in \topmargin=0in \oddsidemargin=0in

\vspace{ -3cm}
\thispagestyle{empty}
\vspace{-1cm}

\

\

\begin{center}
\vspace{1cm}
{\Large\bf

The quantum AdS$_5\times$S$^5$ superstring \\
at finite coupling
}

\end{center}

\vspace{.2cm}



\begin{center}

R.~W.~McKeown and R.~Roiban

\end{center}

\begin{center}
{
\em
\vskip 0.08cm
\vskip 0.08cm
Department of Physics, The Pennsylvania  State University,\\
University Park, PA 16802 , USA
\vskip 0.08cm
\vskip 0.08cm

}
 \end{center}



\vspace{1.5cm}

\vspace{.2cm}

\begin{abstract}

\noindent

The integrability of string theory in  AdS$_5\times$S$^5$  and of the dilatation operator of $\NeqFour$ 
super-Yang-Mills theory has been used to propose an exact solution to the spectral problem in these 
theories. Weak coupling perturbation theory both in gauge theory and on the worldsheet has been 
extensively used to verify this solution.

We discuss worldsheet methods for finding the spectrum of the \adss~ superstring at finite values of the
coupling constant and illustrate them by recovering, within numerical errors, the predictions of the BES 
equation for the universal scaling function. This is the first finite-coupling calculation in this theory which 
uses field theory methods.

\end{abstract}

\newpage

\tableofcontents

\newpage

\section{Introduction}

Superstring theory in \adss~ is described by a complicated interacting two-dimensional field theory of Green-Schwarz type;
solving it exactly appears to be a difficult problem. 
From a conformal field theory perspective one is interested in finding vertex operators labeled by PSU$(2,2|4)$ quantum 
numbers ${\hat {\rm C}}=(E; S_1, S_2; J_1, J_2, J_3)\equiv (E; {\rm C})$, their two-dimensional anomalous dimensions 
$h(\sql,{\hat{\rm C}})$
and their correlation functions. 
The marginality condition, identifying the physical states of the corresponding string theory, determines the energy of the 
state -- or the dimension of the dual ${\cal N}=4$ super-Yang-Mills (sYM) theory operator -- in terms of its charges and 
the 't~Hooft coupling, $E=\Delta=E(\sql,{\rm C})$.
In the static gauge the string energy $E$ has the more direct worldsheet interpretation as the worldsheet energy.

The complete worldsheet theory is classically integrable  \cite{Bena:2003wd}; together with the integrability of the
planar dilatation operator of the dual gauge theory this suggests that the worldsheet theory is integrable at finite
values of the coupling. \footnote{Classical integrability was also shown in the pure spinor formalism in \cite{Vallilo:2003nx}; 
in \cite{Berkovits:2004xu} it was argued that the theory is also quantum-integrable.}
Assuming all-orders integrability Asymptotic Bethe Equations (ABA) \cite{Beisert:2006ez},
Thermodynamic Bethe Ansatz (TBA) equations \cite{Arutyunov:2009ur} and TBA  equations 
in Y-variables \cite{Gromov:2009tv} have
been formulated for the spectrum of target space energies of long and general string states, respectively.

A remarkable feature of the worldsheet fluctuations around the BMN vacuum is that their exact dispersion relation
\be
\epsilon^2 = 1+\frac{\lambda}{\pi^2}\sin^2 \frac{\pi p_\text{ws}}{\sql} 
\ee
bears a distinct similarity with that of a scalar particle on a space with a discrete space-like direction (with spacing 
$a=2\pi/\sql=1/2g$) 
with the important difference that {\it a priori} the momentum $p_\text{ws}$ is not discrete. While this structure is not 
immediately manifest on the worldsheet and its consistency has been checked only in certain 
limits \cite{Hofman:2006xt, Klose:2007rz, Berenstein:2009qd}, it can be derived from various  perspectives in the 
dual gauge theory \cite{Gross:2002su, Santambrogio:2002sb, Beisert:2005tm}. 

The integrability-based predictions for the energies of certain long strings and long operators ({\it i.e.} strings/operators 
carrying at least one large quantum number) have been tested through four loops at weak gauge theory 
coupling~\cite{Bern:2006ew, Cachazo:2006az} and 
through two loops at strong coupling (weak worldsheet coupling expansion) \cite{Giombi:2009gd, Giombi:2010fa}.
For short strings and short operators, the results of TBA/Y-system equations \cite{Leurent:2012ab, Gromov:2009zb} have 
been confirmed through five loops  at weak gauge theory coupling \cite{Eden:2012fe} and through  one loop at strong 
coupling \cite{Roiban:2011fe, Gromov:2011de,Beccaria:2012xm}.
While extensive and quite suggestive that integrability does indeed hold to all orders, such tests cannot definitively 
answer questions like:

\begin{enumerate}

\item
Is it the target space energy operator that is diagonalized by the Bethe ansatz or is it another operator that differs from it
at finite coupling or at sufficiently high order in weak/strong coupling perturbation theory?

\item
Strong coupling perturbation theory is at best an asymptotic series which in certain cases \cite{Basso:2007wd}
is known not to be Borel summable; is it possible to recover the weak gauge theory coupling results for
anomalous dimensions of local gauge-invariant operators from worldsheet calculations?

\item
The worldsheet theory should be a finite quantum field theory; how sensitive are the results obtained in this theory
on the regularization scheme?

\item
How can one find the target space energy of short string states that cannot be (formally) described as limits of classical
string solutions but rather have only a description in terms of vertex operators ({\it e.g.} chargeless states)?


\end{enumerate}

Our aim in the present paper is to initiate the exploration of discrete approaches to the Green-Schwarz string in \adss.  
%
%
Using techniques of lattice field theory\footnote{In the context of the gauge/string duality lattice field theory was used to 
study the Plane Wave Matrix Model \cite{Catterall:2010gf} and the 16-supercharge $0+1$-dimensional sYM theory
\cite{Catterall:2007me}. Two-dimensional sigma models were studies in \cite{Catterall:2006sj}.} 
to evaluate numerically the energy of a particular long string state at finite values of the 
't~Hooft coupling,
in this paper we shall present evidence that the answers to the first two questions above are positive. While we shall not
explicitly address the scheme dependence of the worldsheet calculations, we shall see that our results are consistent 
with the scheme chosen in worldsheet perturbative calculations.
We shall also describe strategies for finding the spectrum of generic string states both in this framework 
as well as in worldsheet perturbation theory.\footnote{Discrete approaches to the worldsheet theory have been proposed 
\cite{Siegel:1994xr} as a systematic means of identifying the perturbation theory of the target space effective field theory from
a worldsheet standpoint. In the context of the AdS/CFT correspondence this approach has been discussed 
in \cite{Hatsuda:2002wf, Nastase:2000za}.}

Lattice field theory \cite{Wilson:1974sk} 
is used extensively to study certain finite-coupling aspects of QCD as well as of condensed matter systems
that can be analyzed in Euclidean setting. One constructs a discrete action on a four-dimensional square lattice which in the 
continuum limit becomes the desired action and evaluates the path integral numerically though Monte Carlo techniques. The 
partition function is arguably the simplest quantity to evaluate. Discrete forms of operators can also be placed on the lattice and 
the corresponding path integral yields their correlation functions.
Here we will follow this strategy, discuss possible square-lattice discretizations of the Green-Schwarz string in \adss~ 
and describe calculations that yield (in principle) quantities of interest for the dual gauge theory. 

The properties of the theory are both a source of simplifications and complications. On the one hand, 
all fields are scalars (albeit some of them are anticommuting\footnote{From this perspective these fields resemble 
the topologically-twisted fermions proposed in \cite{Catterall:2005df} as a way to realize supersymmetric field theories 
on the lattice, see \cite{Catterall:2009it} for a review.}) so their discretization is to some extent straightforward.  
On the other, Grassmann-odd fields
can appear more than quadratically.  Since it is difficult to simulate numerically anticommuting fields, they are 
usually integrated out; here however it is necessary to first linearize the terms that are more than quadratic and 
this potentially leads to a proliferation of auxiliary fields. 
Calculations then proceed by integrating out the Grassmann-odd  fields and exponentiating the resulting determinant 
such that the number of propagating fields in the continuum limit is correct. Absence of anomalies makes this more 
straightforward than in four dimensions. 

To illustrate the discrete approach to the Green-Schwarz string in \adss~ we shall evaluate the energy of the long 
folded string for $1\le g \le 40$ by computing the partition function of the discretized AdS light-cone gauge action 
in the background of the null cusp solution; we reproduce within reasonable accuracy the results of the BES equation.
This is the first finite-coupling calculation in this theory using field theory methods.
We shall use the standard Rational Hybrid Monte Carlo (RHMC) algorithm \cite{Kennedy:1998cu, Clark:2003na, Clark:2006wq} 
for an efficient treatment of the fermion determinant contribution (R) and  for the evaluation of the path integral (HMC). 
%

We begin in section~\ref{numerics} with a discussion on the discretization of the various forms of the Green-Schwarz 
action and brief comments on Monte Carlo methods for the evaluation of path integrals while relegating more details to 
Appendix~\ref{RHMC}. We also outline three possible approaches to determining the worldsheet spectrum and other
interesting quantities from numerical calculations. In section~\ref{explicitcalculation} we discuss in detail the example 
of the calculation of the universal scaling function.  We outline the construction of the discrete version of the AdS 
light-cone action in the null cusp background and point out the discrete derivative required for a stable fermionic 
contribution,  illustrate the evaluation of the partition function for one value of the coupling constant, discuss the 
dominant source of errors in the simulation and present our results. 
In section~\ref{conclusions_and_more} we summarize our conclusions.

\section{A numerical approach to the Green-Schwarz string \label{numerics} }

Lattice field theory provides a means to evaluate observables of the Green-Schwarz string in \adss~
that are accessible on a Euclidean worldsheet, such as energies of string states or dimensions
of the dual gauge-invariant operators, correlation functions of worldsheet operators, expectation
values of Wilson loops, {\it etc.}
In this section we discuss general features of possible approaches to the discretization of the Green-Schwarz action
in \adss~ as well as  review and extend possible approaches to the calculation of target space
energies of string states that may be implemented in this framework.

\subsection{
A first pass at a discrete Green-Schwarz action in \adss
\label{numericsP1}}

To simulate a quantum field theory on a Euclidean lattice one begins with the continuum Euclidean action
and discretizes it in a way that preserves as many of the continuum symmetries as possible.
Lattice fields are assigned to either links (such as gauge fields) or to nodes (such as scalar fields).
Under gauge transformations the link variables transform non-locally, depending on the beginning
and end of the link, {\it e.g.} $L_{n, m}\mapsto U_n L_{n,m}U^\dagger_m$ while the node variables
transform locally.
Further considerations in the construction of the discrete action is the error introduced by the finite
size lattice and the speed of the convergence of the continuum limit, the restoration of broken
symmetries in the continuum limit, the existence of nontrivial renormalization
and in particular of quantum-generated terms proportional to inverse lattice spacing which spoil the properties of the classical
continuum limit, the proper treatment of fermions, the reality of the fermion contribution to the partition
function (known as the fermion sign/phase problem \cite{Giedt:2003vy}), the existence of anomalies and the solution
to the fermion doubling problem {\it etc.}

Many of the issues present in standard matter-coupled gauge theories are absent in the Green-Schwarz
string in curved space \footnote{This is true also in flat space, but that theory is effectively free.}.
For example, from the worldsheet standpoint all fields are scalars and thus they are uniformly assigned to
lattice sites.
Also, for a real bosonic  background the fermionic contribution to the partition function 
is real and consequently there cannot be a phase ambiguity.
Moreover, since the worldsheet theory is expected to be finite (and conformal in an appropriate gauge)
one does not expect that renormalization is necessary and thus no counterterms need to be included in the action;
however, composite operators whose correlation functions we may be interested in computing should 
receive infinite normalization so they should require counterterms.  
Last but not least, no two-dimensional anomalies are present and thus the doubling problem of fields 
with linear quadratic terms can be resolved without resorting to the usual four-dimensional techniques 
(see {\it e.g.} \cite{Wiese:2009qsa} for an introductory review).

The fact that $\kappa$ symmetry is different from a standard gauge symmetry -- in that it acts
nonlinearly and does not have an independent gauge field -- suggests that it must be treated
differently from usual local symmetries.
A possible approach -- which we will adopt here and use in the explicit calculation in
section~\ref{explicitcalculation} --  is to discretize the gauge-fixed action.
Since $\kappa$ symmetry is related to worldsheet supersymmetry such an approach potentially
breaks it. One may however test whether this is the case and, if necessary,
correct for such effects.

There are several (classes of) actions that one might consider as the starting point for the construction
of a lattice action:
\begin{enumerate}

\item
Poincar\'e patch conformal gauge actions are typically simpler  but they require
the presence \cite{Kallosh:1998nx, Kallosh:1998ji, Roiban:2000yy} of an extended string background.
One may in principle avoid this by choosing a light-cone type $\kappa$-symmetry gauge while maintaining
conformal gauge for two-dimensional diffeomorphisms. While the resulting action cannot be used for
perturbative calculations (unless one further chooses a bosonic background) due to the absence of
a free-fermion quadratic term, it may nevertheless be useful for lattice calculations.

\item
Actions in which all constraints are solved are an appealing starting point for a construction of a discrete action
as one needs not worry about the restoration of the corresponding symmetries in the continuum limit. Examples are
the AdS-light-cone gauge~\cite{Metsaev:2000yu} and the uniform light-cone gauge \cite{Arutyunov:2009ga}
actions\footnote{Technical difficulties may arise in the latter case due to the Nambu square-root form of the uniform
light-cone gauge action.}.

\item
One may consider discrete projective light-cone gauge actions already put forward in the literature
\cite{Siegel:1994xr, Nastase:2000za, Hatsuda:2002wf}.

\end{enumerate}

\noindent
The appropriate choice depends on the observable to be computed and on the computational strategy.
In the next subsection we will discuss approaches to the calculation of target space energies that make use
of the first two types of actions and in section~\ref{explicitcalculation} we will use a discretized AdS-light-cone action.

The discretization of the bosonic action is straightforward: since all fields are scalars they are assigned to lattice sites
\be
\phi(\tau, \sigma)\mapsto \phi_{(m,n)}
\ee
and their derivative in the direction of a unit two-dimensional vector ${\vec v}$ is replaced by a finite difference
\be
\partial_{{\vec v}} \phi \mapsto \frac{1}{a}(\phi_{(m,n)+{\vec v}} -\phi_{(m,n)}) \ ,
\label{discrete_derivative}
\ee
where $a$ is the lattice spacing. The error introduced by this replacement is ${\cal O}(a)$;
depending on the desired precision for the calculation more involved discrete approximations for the continuum derivatives
may be necessary. In Appendix \ref{9ptstencil} we include details on a nine-point approximation used in the calculation in
section~\ref{explicitcalculation}.

A common treatment of Grassmann variables is to integrate them out analytically; if higher-point fermion interactions
are present they are first linearized by introducing  an appropriate set of auxiliary fields auxiliary fields. 
%
%
The resulting determinant is either evaluated directly or exponentiated in terms of commuting pseudo-fermions,
\be
\int D\Psi e^{-\int d^2\xi {\psi}M\psi} = (\det M)^{1/2}=(\det MM^\dagger )^{1/4} =\int D \zeta D {\bar \zeta}\; 
e^{-\int d^2\xi \, {\bar\zeta} (MM^\dagger)^{-1/4} \zeta } \ .
\label{fermiondet}
\ee
In general, care must be taken to avoid double-counting the fermionic contribution in the continuum limit. In our case 
one can show that the Pfaffian of $M$ is real and thus it
is sufficient to exponentiate the operator $MM^\dagger$; since it contains a Klein-Gordon operator (in general up to a
field-dependent prefactor which is expected to effectively acquire a vacuum expectation value), its discretization is similar
to that of regular commuting scalars  and is free of unwanted doublers \cite{Banks:1979yr}.

The strategy for computing correlation functions in lattice field theory is to approximate the path integral 
in terms of finitely many field configurations which sample the entire phase space.  A  very efficient algorithm for generating 
these field configurations is the RHMC algorithm \cite{Kennedy:1998cu, Clark:2003na, Clark:2006wq}, which we shall review 
in some detail in Appendix~\ref{RHMC}. With a specific rational 
approximation for the fractional power of the fermion matrix \eqref{fermiondet}, one defines an evolution Hamiltonian 
\footnote{This Hamiltonian is the evolution operator along some fictitious (Monte Carlo) ``time" direction $\tau$.}
by adding to the discrete action the squared conjugate momentum for each field, randomly generates some field configuration 
and then generates further ones by repeating the following steps:
\begin{enumerate}

\item randomly generates momenta  for all fields from a Gaussian distribution

\item evolves the field configuration with the evolution Hamiltonian (Molecular dynamics) for some ``time" interval $\Delta\tau$

\item accepts the initial or the final field configuration stochastically, though a Metropolis acceptance test (which, in some 
sense, ``decides" whether the final field configuration can be the result of a quantum mechanical evolution).  
This step eliminates errors introduced at step $2)$ due to various approximations.

\end{enumerate}

\noindent
It has been shown in \cite{Kennedy:1998cu} that this algorithm leads to field configurations which cover the entire phase 
space of the system and can be used to construct the partition function or correlation functions of operators.

\subsection{Various approaches to energy calculations}

The semiclassical expansion has been extensively used to study the worldsheet perturbative expansion of energies of long
strings in \adss~ -- {\it i.e.} strings with large quantum numbers  and thus dual to ÒlongÓ sYM operators with large canonical
dimensions (see {\it e.g.}  \cite{Beisert:2010jr} for a review). It was suggested in \cite{Tirziu:2008fk} that similar techniques
may also be applied to short strings, provided that the corresponding state can be obtained by analytic continuation
from a long string state. A strategy that uses the conformal dimension of worldsheet vertex operators and potentially yields
the energy of generic string states,  has been discussed in \cite{Roiban:2009aa}. Here we briefly review these techniques
and add another one based on the worldsheet construction of two-point functions of local gauge theory operators and phrase
them such that they are amenable to numerical calculations based on discretized worldsheet actions in \adss.

\subsubsection{A conformal gauge approach}

In conformal gauge the worldsheet theory for strings in \adss~ is a conformal field theory -- albeit one which is neither
factorizable nor particularly suited to a perturbative treatment due to the special features of the Green-Schwarz fermions.
Conformal invariance determines the form of the two-point function of local worldsheet operators; using it we may extract
target space information from a (direct numerical) evaluation of the two-point function of fairly general local worldsheet
operators.

Indeed, a general local operator $W(\xi)$ can be expanded in the basis of local operators  with definite worldsheet
dimension $h_n$ as
\be
W(\xi)=\sum_n c_{W, n} \V_n(\xi) \ .
\label{VopDec}
\ee
The general form of the two-point function of the basis elements
\be
\langle \V_m(\xi_1) \V_n(\xi_2) \rangle = \delta_{m,n} \frac{c_V}{|\xi_1-\xi_2|^{2 h}}
\ee
implies that the two-point function $\langle W(\xi_1) W(\xi_2) \rangle$ is
\be
\langle W(\xi_1) W(\xi_2) \rangle =\sum_n \frac{c_n c^2_{W, n}}{|\xi_1-\xi_2|^{2 h_n}}  \ ,
\label{ws2pf}
\ee
where the normalization factor $c_V$ and the conformal dimensions $h_n$ depend on the
worldsheet coupling constant and the charges of the operator $\V_n$ under various target
space symmetries.
Assuming some way of determining the worldsheet two-point function of two operators $W(\xi)$, the lowest
worldsheet conformal dimension that appears in \eqref{ws2pf} may be extracted from the large distance asymptotics.
More generally, given a sufficiently precise determination of $\langle W(\xi_1) W(\xi_2) \rangle$, the dimensions $h_n$ 
of other vertex operators that enter the decomposition \eqref{VopDec} may in principle be extracted through a Fourier 
analysis; the coefficients $c_n$ and $c_{W, n}$ need not be known.

Then, the condition that $\V_n$ are exactly marginal,
\be
h_n = 2 \ ,
\label{physical state}
\ee
determines the target space energy of the corresponding string state in terms of its charges and the 't~Hooft
coupling.

To carry out such a program it is useful to choose an operator $W$ with definite charge under the target
space symmetries which also depend explicitly on the target space energy or boundary conformal dimension; 
the classical bosonic vertex operators discussed in conformal gauge in \cite{Polyakov:2001af, Tseytlin:2003ac, 
Roiban:2009aa} are possible candidates.
As discussed in section~\ref{numerics}, from the perspective of a numerical calculation of the two-point function
\eqref{ws2pf}, purely bosonic approximate vertex operators are very useful because fermions can be integrated out
analytically.
Bosonic vertex operators are naturally constructed in terms of the embedding coordinates
\be
&&
 Y_a  Y^a = Y_+Y_+^* - Y_x Y_x^*  - Y_y Y_y^*  =1  , \ \ \ \ \ \
 Z_k Z_k = Z_x Z_x^*  + Z_y Z_y^* + Z_z Z_z^*=1,   \label{nen}
\ee
where
$  Y_+ = Y_0 + i Y_5, \   Y_x= Y_1 + i Y_2, \  Y_y= Y_3 + i Y_4,$ \  $
Z_x=Z_1 + i Z_2, \  Z_x=Z_3 + i Z_4, \ Z_z=Z_5 + i Z_6$ and the relation to
the Poincar\'e patch is the usual one
\be
Y_m=\frac{x_m}{z}\,, \quad
Y_4=\frac{1}{2 z}(-1+z^2 +x^m x_m) \,, \quad
Y_5 = \frac{1}{2 z} (1+z^2 +x^m x_m) \ .
\ee
In the embedding coordinates the conformal gauge bosonic action is
\be
S= { \sql \ov 4 \pi}  \int d^2 \s  \ \Big(  -\partial Y_a \cdot\partial Y^a +  \partial Z_k \cdot\partial Z_k
+ {\rm fermions\ } \Big)  \ .
 \ee
To describe vertex operators it is useful to do a Euclidean rotation both on the worldsheet and in the
target space
\be \la{p}
t_e=i t  \ , \ \ \ \ \ \    Y_{0e}= i Y_0 \ , \ \ \ \ \ \ x_{0e} = i x_0 \ ,
\ee
so that $ Y^M Y_M= -Y_5^2 +Y_{0e}^2   + Y_i Y_i  +Y_4^2 =-1$. Then, unintegrated vertex operators have
the form
\be
\V \sim (\rY_+)^{-\D} \ \Big[ (\del^s Y)^r ... (\bd^m Z)^n + ... \Big]
\equiv (\rY_+)^{-\D}\ U (Y,Z,...)
\label{unintegrated_bosonic} \ .
\ee
where
\be
   K(x,z) =k_\Delta\ ( \rY_+)^{-\D}
              = k_\Delta\ \left(z+ z^{-1} x^m x_m\right)^{-\Delta}~~, \ \ \ \ \text{and} \ \ \ \
	\rY_+ \equiv Y_5 + Y_{4}
 \la{gh}
 \ee
is the usual bulk-to-boundary propagator in AdS space.
\footnote{As explained in \cite{Polyakov:2001af, Tseytlin:2003ac}, the structure of the
vertex operator follows closely that of flat space vertex operators, with $ (\rY_+)^{-\D}$ being the Euclidean analog of $e^{-iEt}$
and $U (Y,X,...)$ carrying the dependence on the string level, symmetry transformations, {\it etc.}}

On a discrete worldsheet the unintegrated vertex operator is placed at a lattice site
\be
\V(\xi) \mapsto \V_{(m,n)}
\ee
and depends on as many adjacent sites as are used to define the discrete derivative of a field (the number
of such derivatives appearing in $\V$ is related to the string level). The two-point function of such operators (specified by
the parameter $\Delta$ and a choice of $U(Y, Z, \dots)$)
\be
\langle  \V_{(m_1,n_1)}\V_{(m_2,n_2)}\rangle =\frac{1}{Z}\int D[x_{(m,n)}, z_{(m,n)}^M, \theta_{(m,n)}, \eta_{(m,n)}]\;
e^{-S_\text{discrete}} \V_{(m_1,n_1)}\V_{(m_2,n_2)}
\ee
may then be computed through Monte Carlo techniques which we will review in Appendix~\ref{RHMC}
as a function of $\Delta$, the symmetry charges, the worldsheet coupling and the lattice separation
$d_{12}^2=((m_1-m_2)^2+(n_1-n_2)^2)$. \footnote{In computing this correlation function it may be necessary 
to add to $\V_{(m,n)}$ countertems proportional to the inverse lattice spacing; this is due to the fact that 
in the continuum theory operators are expected to receive infinite renormalization responsible for their 
two-dimensional anomalous dimension.\label{operator_renormalization}}
Extracting from it the worldsheet conformal dimensions $h_n$ (or perhaps only the smallest one)\footnote{It is
possible to argue that, in the context of our discussion, one can unambiguously follow one worldsheet dimension from weak
to strong worldsheet coupling.  Indeed, it was suggested in \cite{Polyakov:2001af} that the dimensions of gauge theory
operators obey a non-intersection principle -- {\it i.e.} that there should be no level crossings for states with the same
quantum numbers as $\lambda$ changes from small to large values. Then, the general structure of the worldsheet
anomalous dimension  \cite{Tseytlin:2003ac, Roiban:2009aa}
$$
h_n=\Delta_n(\Delta_n-c) + F_n(\sqrt{\lambda}, Q) \ ,
$$
together with the positivity of $\Delta$ and $h$ imply that the worldsheet (anomalous) dimensions  should also obey a non-intersection principle.} as a function of $\Delta$,
$\lambda$ and identifying the curve with the equation \eqref{physical state} yields the target space energy
$\Delta=\Delta(\lambda, \dots)$.

Since the discrete action is not conformally invariant, the extraction of the worldsheet conformal dimensions $h_n$
requires a careful consideration of the continuum limit. \footnote{It is possible that the leading term in the large distance
expansion is less sensitive to the details of the discretization and may be used to extract the smallest worldsheet dimension
even on a finite-size lattice.
}
In turn, the curve with equation \eqref{physical state} yields the target space energy  $\Delta=\Delta(\lambda, \dots)$.

By not imposing the Virasoro constraint from the outset as well as by not focusing from the outset on target space quantities,
the outcome of this algorithm potentially provides us with a wealth of information on conformal field theories of Green-Schwarz
type. For the goal of finding the energies of string states these features make it however computationally intensive
due to the large number of intermediate steps necessary to find $\Delta$ from the worldsheet two-point function.
%
It is therefore important to identify strategies that require fewer/simpler auxiliary quantities.

\subsubsection{Operator dimensions from the boundary two-point functions}

Integrated vertex operators  \cite{Tseytlin:2003ac, Roiban:2010fe}, labeled by a point on the boundary of AdS space,
are dual to local gauge-invariant operators of the boundary theory. They provide a means to construct the two-point
function of general local operators in the boundary theory from which one may extract directly the eigenvalues of the
dilatation operator.


A general local gauge-invariant operator ${\rm O}(\rx)$ in the boundary theory can always be decomposed in a basis of
local operators ${\cal O}_n(\rx)$ with definite (anomalous) dimensions
\be
{\rm O}(\rx)=\sum_n\; c_{{\rm O},n}(\lambda){\cal O}_n(\rx) \ .
\label{decomposition}
\ee
Using the form of the two-point functions of ${\cal O}_n(\rx)$ dictated by conformal invariance,
\be
\langle {\cal O}(\rx){\cal O}(\ry)\rangle = \frac{c_{\cal O}(\lambda)}{|\rx-\ry|^{2\Delta_{\cal O}}}\ ,
\ee
($c_{\cal O}(\lambda)$ is a potentially nontrivial normalization factor) it follows that the two-point function of
the operators ${\rm O}(\rx)$ is
\be
\langle {\rm O}(\rx){\rm O}(\ry) \rangle = \sum_n\frac{c_n(\lambda) c_{{\rm O},n}^2(\lambda)}{|\rx-\ry|^{2\Delta_n}} \ .
\label{expansion}
\ee
Thus, assuming that it is possible to find the exact two-point function of ${\rm O}(\rx)$,
it is then possible to extract the dimensions of the operator of lowest dimension in its spectral
decomposition \eqref{decomposition} from the large-distance behavior of the two-point
function \eqref{expansion}; the dimensions of the other operators in \eqref{decomposition} may also be
extracted through a Mellin transform or by simply fitting the two-point function to an expression of the type \eqref{expansion}.

It is in general difficult to identify the vertex operator corresponding to a specified gauge theory operator; to carry out the 
program above however it suffices to know the two-point function of {\it some} operator ${\rm O}$. Thus, one can simply pick 
any desired integrated vertex operator $V(\rx)$ and, if its two-point function can be computed, it can in principle be used to 
extract the dimension of the gauge theory operators with nonvanishing overlap with its dual ${\rm O}_V$. 
It is intuitively clear that, to extract information from a numerical (and thus inherently not exact) evaluation of 
the two-point function $\langle V(\rx) V(\ry)\rangle$, it is useful to have as few dominant terms as possible 
on the right-hand side of eq.~\eqref{expansion}. 
One may, of course, target operators with a specific PSU$(2,2|4)$ quantum numbers ${\rm C}=(S_1,S_2;J_1,J_2,J_3)$
by choosing a worldsheet operator with the same charges. \footnote{Operators with identical charge vectors ${\rm C}$ may be 
further distinguished by hidden local charges such as those related to the integrable structure of the planar theory. }
Further super-selection sectors are introduced by the gauge theory engineering dimension and the string level. At least at 
small and large 't~Hooft coupling they guarantee that the dimensions $\Delta$ differ by ${\cal O}(1)$ and 
${\cal O}(\lambda^{1/4})$ quantities respectively and thus the features of the two-point function are expected to be well-separated
in Mellin space.

The integrated vertex operator associated to the unintegrated vertex operator \eqref{unintegrated_bosonic} is
\cite{Tseytlin:2003ac, Roiban:2010fe}
\be
 V(\rx)= \int d^2 \xi \  \V \big( x(\xi) -\rx;\ ... \big) =  \int d^2 \xi \  [K(x(\xi)-\rx,z(\xi))]^{-\Delta}
  \ U[x (\xi)-\rx, Z(\xi)]  \ ;
 \label{integrated_bosonic}
\ee
Its form on the discrete worldsheet is obtained by using the discrete fields and replacing the integral
with a sum over all lattice sites,
\be
 V(\rx)=\frac{1}{a^2}\sum_{m,n}\V_{(m,n)} \big( x_{(m,n)} -\rx;\ ... \big) \ .
\ee
The boundary two-point function of the dual operator ${\rm O}_V(\ry)$ is then given by
\be
\langle {\rm O}_V(\rx){\rm O}_V(\ry) \rangle=\langle  V(\rx)V(\ry)\rangle
=\frac{1}{Z}\int D[x_{(m,n)}, z_{(m,n)}^M, \theta_{(m,n)}, \eta_{(m,n)}]\;e^{-S_\text{discrete}} V(\rx)V(\ry)
\ee
and, as in the case of the two-point function in the previous section, may then be computed through Monte Carlo techniques.
\footnote{The comments as in footnote~\ref{operator_renormalization} apply to this calculation as well.}

For each value of the 't~Hooft coupling this algorithm requires the evaluation of a single auxiliary function of one 
variable -- the boundary separation of operators  $|\rx_1-\rx_2|$  --
and the dimensions of operators in the boundary theory follow directly from it.
This two-point function may be computed either with a discretized conformal gauge action or with a discretized
action in a physical gauge, such as the light-cone gauge. In the latter case however it may be useful to choose
a slightly different worldsheet operator $V(\rx)$ than described above.

Indeed, in this gauge one of the Cartan generators of the boundary rotation group is spontaneously broken and thus it
cannot be used to label $V$. Moreover, the worldsheet field $x^-$ is nonlocal (with local derivatives) and thus not 
easy to use as an argument  of $\V$ in eq.~\eqref{integrated_bosonic}. A straightforward resolution is to simply not 
include the bulk-to-boundary factor $K$ in $\V$ and choose a function $U$ which depends on $x^\pm$ only through 
their derivatives:
\be
 V_{\text{lc}}(\rx_\perp)=\int d^2 \xi \   U[x_\perp (\xi)-\rx_\perp, Z(\xi)]  \rightarrow
 \sum_{m,n}U_{(m,n)} \big( x_{\perp, (m,n)} -\rx_\perp;\ ... \big) \ .
 \label{integrated_bosonic_modified}
\ee
While this cannot be called ``vertex operator", it is nevertheless dual to some local boundary operator labeled by the
two coordinates $\rx_\perp$ transverse to the light-cone. Interpreting the absence of $\rx^\pm$ as they having been
set to zero, the dimensions of operators in the boundary theory may be extracted from an expansion analogous to
eq.~\eqref{expansion} which now depends only on $|\rx_{\perp,1}-\rx_{\perp 2}|$.

\subsubsection{Target space energy from a partition function}

The time-honored approach to perturbative calculations of  worldsheet quantum corrections to the
energy of long string states is as the worldsheet vacuum energy in the background of the classical solution
describing the state \cite{Frolov:2002av} (see \cite{Beisert:2010jr} for a review and a complete list of references).
It has been argued in \cite{Frolov:2006qe} and further elaborated on in \cite{Roiban:2007ju, Giombi:2010fa} that,
for single-charge string states, this is also the worldsheet free energy.  More precisely, for single-charge
solutions for which the AdS global time is related to the worldsheet time as $t=\kappa \tau$ the target space
energy is also given by
\be
E = - \frac{1}{\kappa} \ln Z \ .
\label{EofZ}
\ee
More involved expressions relate the partition function and the energy of multi-charge states for which all
parameters may be interpreted as chemical potentials for various charges \cite{Roiban:2007ju, Giombi:2010fa}.
It was moreover argued in \cite{Roiban:2009aa, Roiban:2011fe, Gromov:2011de, Beccaria:2012xm} that similar
semiclassical techniques capture correctly the energy of  short string states which can be interpreted as the small
charge continuation of long string states.

The evaluation of the partition function of a theory is arguably the simplest lattice field theory calculation and thus
provides a good testing ground for the applicability of this framework to the Green-Schwarz string.
Thus, a strategy -- which we will illustrate in section~\ref{explicitcalculation} -- is to
first find the complete continuum action of fluctuations $\tilde\Phi$ around the desired classical solution
$\Phi_\text{cl}$ and discretize it as discussed in sec.~\ref{numericsP1}. \footnote{ {\it A priori} this action may exhibit
position dependence inherited from the classical background (if the classical solution is inhomogeneous).
This simply translates into an explicit dependence of the Lagrangian on the lattice site (and on the lattice
spacing) apart from that of fields.} 
As before, the action $S_{E, \text{discrete}}$ is the sum over all lattice sites of the discrete Lagrangian.
Then we evaluate the partition function
\be
Z=\int D\Phi_{(m,n)} e^{-S_{E, \text{discrete}}}
\ee
as a function of the parameters of the solution ({\it i.e.} for many choices of those parameters and then construct an
interpolating function) and the 't~Hooft coupling; from it one extracts the energy of the corresponding string state either though
\eqref{EofZ} or  though the more involved relations derived in  \cite{Roiban:2007ju, Giombi:2010fa}.

This approach should be insensitive to the details (such as gauge choices) of the continuum action. However, only long
string states whose Euclidean action is real can be considered; this is due to the fact that, in the Monte Carlo evaluation
of path integrals, the corresponding probability measure, $\exp(-S_E)$, is required to be real.
Many interesting solutions -- such as the spin-$S$ folded string (with and without angular momentum on S$^5$),
various circular string solutions whose analytic continuation to small charges describe members of the Konishi multiplet,
{\it etc.} --  satisfy these restrictions. In the next section we discuss in detail an application of this strategy.

\section{  An example: the universal scaling function at finite coupling   \label{explicitcalculation}}

The universal scaling function is expected to be the solution to the BES equation; this equation reproduces
the first few orders in the weak \cite{Bern:2006ew, Cachazo:2006az} and strong  \cite{Frolov:2006qe, Giombi:2009gd} 
coupling expansions that have been computed directly:
\be
f(\lambda)\big|_{\lambda\rightarrow 0}&=&
8g^2\left[1-\frac{\pi^2}{3}g^2+\frac{11\pi^4}{45}g^4-\left(\frac{73}{315}+8 \zeta_3\right)g^6+\dots\right]
\\
f(\lambda)\big|_{\lambda\rightarrow\infty}&=&4g\left[1- \frac{3\ln 2}{4\pi g}- \frac{\rm K}{16\pi^2 g^2}+...\right] 
\quad, \qquad
g=\frac{\sql}{4\pi}
\ .
\ee
Further terms in these expansions as well as the value of the universal scaling function at finite values of the coupling
can be obtained from the BES equation\footnote{While strong coupling perturbation theory is not summable, the first 
three terms in this expansion are a good numerical approximation 
to the exact function for $g\ge 1$}. We will reproduce these values (within numerical errors) at
various values of $g\in[1,40]$ by numerically evaluating the worldsheet partition function in the
background of the null cusp  solution. Smaller values of $g$, inside the radius of convergence of sYM perturbation 
theory are, in principle accessible but are currently limited by the accuracy of our simulation.

We take the continuum worldsheet theory to be given by  AdS-light-cone gauge action Wick-rotated to a Euclidean 
worldsheet~\cite{Metsaev:2000yu, Giombi:2009gd}:
\be
S &=& g\int dt \int_0^\infty d s \, {\cal L}_E \quad, 
\\
\mathcal{L}_E &=& \dot{x}^* \dot{x} + (\dot z^M  + \mathrm{i}  z^{-2} z_N
\eta_i {\rho^{MN}}^i{}_j \eta^j)^2  + \mathrm{i}  (\theta^i \dot{\theta}_i +
       \eta^i\dot{\eta}_i - h.c.) -  z^{-2} (\eta^2)^2 \nonumber \\
  && \quad  +  z^{-4} ( x'^*x'  + {z'}^M {z'}^M) + 2 \mathrm{i} \Big[\
       z^{-3}\eta^i \rho_{ij}^M z^M (\theta'^j - \mathrm{i}
       z^{-1} \eta^j  x') + h.c.\Big]\;.
\label{LCADS_euc}
\ee
It has manifest $U(1)\times SO(6)\simeq U(1)\times SU(4)$ symmetry.  The fermions are
complex $\th^i = (\th_i)^\dagger,$ $\eta^i = (\eta_i)^\dagger\ $ $(i=1,2,3,4)$ and transform
in fundamental representation of $SU(4)$; the matrices $\rho^{M}_{ij} $ are the off-diagonal
blocks of six-dimensional gamma matrices in chiral representation and
$(\rho^{MN})_i^{\hphantom{i} j} = (\rho^{[M} \rho^{\dagger N]})_i^{\hphantom{i} j}$ and
$(\rho^{MN})^i_{\hphantom{i} j} = ( \rho^{\dagger [M} \rho^{N]})^i_{\hphantom{i} j}$ are  the
$SO(6)$ generators. The fields $z^M$ are neutral under U(1), $\theta^i$ and $\eta^i$ have opposite
charges and the charge of $\eta_i$ is half the charge of $x$.

\subsection{The null cusp fluctuation action, discretization, and some numerical details \label{action_and_stuff}}

The classical solution of \eqref{LCADS_euc} dual to the null cusp and the action for fluctuations around it
were constructed in  \cite{Giombi:2009gd}  (${\tilde z}$ is the norm of the six-component fluctuation vector ${\tilde z}^M$):
\be
{\cal L}_\text{cusp} &=& {| \partial_t \tilde{x} +\frac{1}{2}\tilde{x} |}^2 + \frac{1}{{\tilde z}^4}{| \partial_s \tilde{x} 
-\frac{1}{2}\tilde{x} |}^2\cr
&+& (\partial_t \tilde{z}^M + \frac{1}{2}\tilde{z}^M + \frac{i}{{\tilde z}^2} \tilde{\eta}_i{(\rho^{MN})^i}_j \tilde{\eta}^j \tilde{z}_N)^2 
+ \frac{1}{{\tilde z}^4} (\partial_s \tilde{z}^M -\frac{1}{2}\tilde{z}^M)^2 \cr
&+& i(\tilde{\theta}^i \partial_t \tilde{\theta}_i + \tilde{\eta}^i \partial_t \tilde{\eta}_i 
+ \tilde{\theta}_i \partial_t \tilde{\theta}^i + \tilde{\eta}_i \partial_t \tilde{\eta}^i)-\frac{1}{{\tilde z}^2}(\eta ^2)^2
\label{WLaction}
\\
&+& 2i\big[\,\frac{1}{{\tilde z}^3}\tilde{\eta}^i(\rho ^M)_{ij}\tilde{z}^M(\partial_s \tilde{\theta}^j-\frac{1}{2}\tilde{\theta}^j
-\frac{i}{{\tilde z}}\tilde{\eta}^j(\partial_s x -\frac{1}{2} x))\cr
&& ~~
 +\frac{1}{{\tilde z}^3}\tilde{\eta}_i (\rho^{\dagger}_M)^{ij}\tilde{z}^M(\partial_s \tilde{\theta}_j-\frac{1}{2}\tilde{\theta}_j
 +\frac{i}{{\tilde z}}\tilde{\eta}_j(\partial_s x^* -\frac{1}{2} x^*))\big] \ .
\nonumber
\ee
As discussed there, the universal scaling function is proportional to the worldsheet free energy
\be
Z_\text{string}=e^{-W(g)}\quad, \qquad  W(g) = \frac{1}{2}f(\lambda){\cal V}  =  \frac{1}{8}f(\lambda)
\int dt \int ds \; 1 ~ \ .
\label{free_energy}
\ee
The various numerical factors are related to the coordinate transformation and field redefinition between the long 
folded (GKP) string in global AdS coordinates and the null cusp solution in the Poincar\'e patch in light-cone 
gauge \cite{Giombi:2009gd}.

To integrate out the fermions  the quartic terms are linearized by introducing a scalar and an $SO(6)$ vector auxiliary
fields:
\be
-\frac{1}{{\tilde z}^2} ({\tilde \eta}^2)^2 \mapsto \frac{1}{2}{\tilde\phi}^2 + \frac{\sqrt{2}}{{\tilde z}}{\tilde\phi} {\tilde\eta}^2
\quad,\quad
 -\frac{1}{{\tilde z}^{4}}({\tilde z}_N\,{\tilde \eta}\rho^{MN}{\tilde \eta})^2 \mapsto  \frac{1}{2}({\tilde\phi}_M)^2
+ \frac{\sqrt{2}}{{\tilde z}^{2}}{\tilde\phi}_M{\tilde z}_N\,{\tilde \eta}\rho^{MN}{\tilde \eta} \ .
\ee
The resulting quadratic fermion matrix $M_\text{cusp}$ can be read without difficulty; for completeness we include
it in appendix~\ref{moreonfermions}. Integrating out the fermions and exponentiating the resulting determinant
leads to the action
\be
{\cal L} &=& {| \partial_t \tilde{x} +\frac{1}{2}\tilde{x} |}^2 + \frac{1}{{\tilde z}^4}{| \partial_s \tilde{x} -\frac{1}{2}\tilde{x} |}^2
+ (\partial_t \tilde{z}^M + \frac{1}{2}\tilde{z}^M )^2 + \frac{1}{{\tilde z}^4} (\partial_s \tilde{z}^M -\frac{1}{2}\tilde{z}^M)^2
\cr
&+&\frac{1}{2}{\tilde\phi}^2 +\frac{1}{2}({\tilde\phi}_M)^2+\zeta^\dagger (M^\dagger_\text{cusp} M_\text{cusp})^{-1/4}\zeta \ ;
\label{final_continuum_L}
\ee
it  contains 15 real bosonic fields (8 physical and 7 auxiliary) as well as 16 more complex bosons $\zeta$ 
used to exponentiate the fermion determinant, cf. eq.~\eqref{fermiondet} \footnote{In that
equation the formal variable $\psi$ stands for $\psi\equiv ({\tilde\theta}^i, {\tilde \theta}_i,
{\tilde\eta}^i, {\tilde\eta}_i)$ with $i=1,\dots,4$.}.

Its discretization, as reviewed in section~\ref{numericsP1}, is relatively straightforward. 
Using the block structure of $M_\text{cusp}$ \eqref{fermion-K} one can
easily convince oneself that $M^\dagger_\text{cusp} M_\text{cusp}$ contains the Klein-Gordon operator
as a free limit. As in the case of bosons its discrete version converges to the right continuum limit and is free of unwanted
doublers.  Moreover, its determinant is real and positive and thus its fourth-root is unambiguous. As described in
Appendix~\ref{RHMC}, we will approximate $(M^\dagger_\text{cusp} M_\text{cusp})^{-1/4}$ as a rational function
of $(M^\dagger_\text{cusp} M_\text{cusp})$ \cite{Remez}:
\be
(M^{\dagger}M)^{-\frac{1}{4}} = \alpha_0+\sum_{i=1}^{P}\frac{\alpha_i}
{M^{\dagger}M+\beta_i} \ .
\label{rational_approx0}
\ee
The standard choice $P=15$ leads in our case to an error ${\cal O}(10^{-5})$ for $g\in (10^{-7}, 10^3)$.

In the discretization of the Lagrangian \eqref{final_continuum_L} we could, in principle, use the simplest two-site version of the
discrete derivative \eqref{discrete_derivative}, whose error is ${\cal O}(a)$. It turns out however that, for lattice sizes accessible
to us this approximation is too coarse and leads to numerically-unstable fermion contributions. We found that stability 
is achieved only for an error of ${\cal O}(a^8)$. We must therefore use a nine point stencil for the derivative
\be
a\partial_xf(x) &=& \frac{4}{5}(f(x+a)-f(x-a))- \frac{1}{5}(f(x+2a)-f(x-2a))
\cr
&&\quad
+ \frac{4}{105} (f(x+3a)-f(x-3a))- \frac{1}{280} (f(x+4a)-f(x-4a))+O(a^9) \ .~~~~~~
\label{discrete_derivative_9pt}
\ee
For completeness we include its derivation in Appendix~\ref{9ptstencil}. It is also trivial to choose a discretization that 
preserves the manifest global $SO(6)$ symmetry of \eqref{final_continuum_L}.

To construct the action \eqref{final_continuum_L} from  the action in eq.~\eqref{WLaction} we introduced a number of
Lagrange multiplier fields; to recover the original action one is to use the saddle-point approximation to integrate them
out. While this is harmless in the continuum theory, strictly speaking the partition function of the original theory is recovered
only up to factors of the determinant of the unit operator.
While these factors are expected to be unity in the continuum theory, in the presence of a regulator they may be nontrivial,
albeit coupling constant independent.
Since the initial partition function is such that it equals unity if the action were zero, we will eliminate the potential extra 
factors by dividing by the partition function with $S_E=0$.  We have checked that the corresponding
subtraction term in the free energy decreases as the lattice spacing is decreased, consistent with it being an artifact
of the discretization.

Even though the original action \eqref{final_continuum_L} is defined on an infinite worldsheet, simulating it on
a lattice requires placing it in finite volume. It has been suggested in \cite{Giombi:2010zi} that the finite-size effects 
due to  placing the theory on a cylinder of length $L$ translate into $\delta f\sim  {1}/{L^2}$ corrections to the universal scaling 
function (and to $1/L=1/\ln S$ corrections to the energy of the long folded string). It is easy to argue that 
such effects are of the same order as the minimal finite-volume error (if the lattice has equal sides) and thus cannot be 
extracted from a calculation of the type we will describe here. 

To estimate a lower bound on the finite-volume effects vis-\`a-vis the fact that we are interested in the 
value of the (free) energy we use the uncertainly relations. Taking the time uncertainly to be the same as the length 
of the (Euclidian) time direction of our lattice, $\Delta t=T$,  the uncertainty in the energy is
\be
 \Delta E \ge \frac{1}{2 T} \ .
\ee
By dividing out the length of the time direction in \eqref{free_energy}, the uncertainly in the energy is related to
the uncertainty in the universal scaling function as
\be
\Delta E = \frac{V_2}{8T}\, \Delta f(g) \ ;
\ee
It thus follows that the error on the universal scaling function is bounded from below by
\be
\Delta f(g) \ge \frac{4}{V_2} \ .
\label{lower_bound0}
\ee
Other sources -- such as statistical, discretization, etc -- add to this estimate. If the length of the space-like and Euclidian 
time-like direction are of the same order this error is of the same order as the expected finite volume correction to the 
universal scaling function, $\delta f\sim 1/L^2$. One may attempt to distinguish them by considering an asymmetric lattice
with the time-like direction much larger than the space-like direction. We shall not pursue this here.


\subsection{The simulation, data analysis and results}

To simulate the discretized action \eqref{final_continuum_L} on a lattice we employed the RHMC algorithm reviewed in
appendix~\ref{RHMC}. We used\footnote{The number of lattice sites and the lattice volume are chosen such that 
the simulation runs sufficiently fast while still having reasonably small discretization errors. } 
$10\times 10$ and $12\times 12$ lattices with volume $V_2=\pi^2$ and evaluated
the worldsheet free energy $W$ and the universal scaling function $f(g)$ (cf. eq.~\eqref{free_energy}) for several
values of the coupling
\be
g\in\{1, 2, 5, 10, 15, 20, 30, 40\} \ .
\label{gvalues}
\ee
To avoid potential problems with constant spinors in the regime when fermions are effectively light\footnote{This occurs at 
small values of the 't~Hooft coupling: the free momentum space action looks like $S\sim g(p^2+m^2)$; for fixed $gp^2$ 
the mass term is irrelevant at small values of $g$.} we will use anti-periodic boundary conditions for fermions
while all bosons are taken to be periodic. The fermion boundary conditions are captured by the detailed
structure of the matrix $(M^\dagger_\text{cusp} M_\text{cusp})$ in the discretized theory and arises from the derivatives 
on fermions at the edges of the lattice. It is interesting to note that if the fermion boundary conditions are chosen to be 
periodic the simulation does not appear to converge.

\begin{figure}[t]
  \centering
  \subfigure[]{%
    \includegraphics[scale=.48]{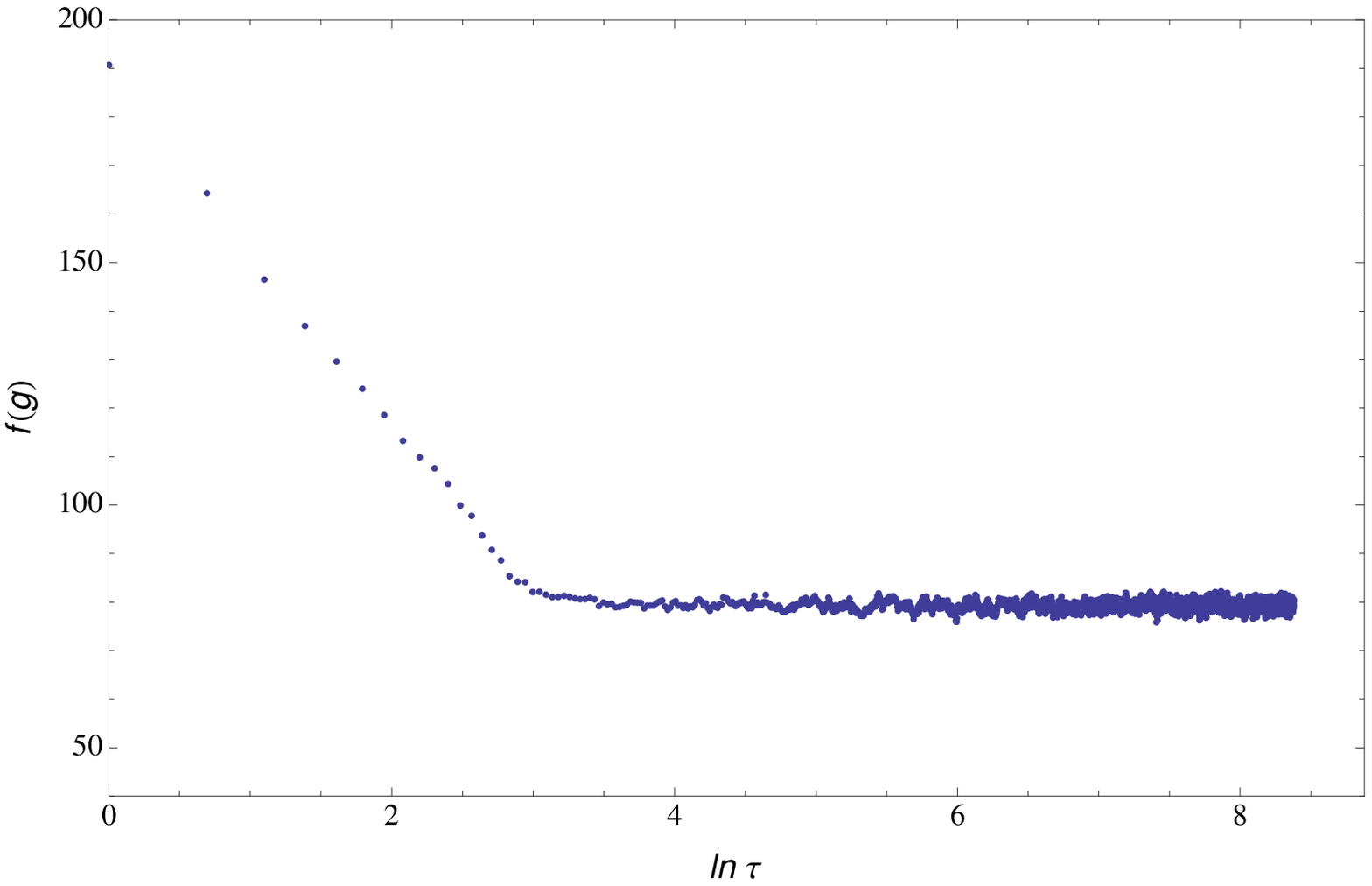}
    \label{thermalization}
  }
  \subfigure[]{%
     \raisebox{-3.5mm}{ \includegraphics[scale=.32]{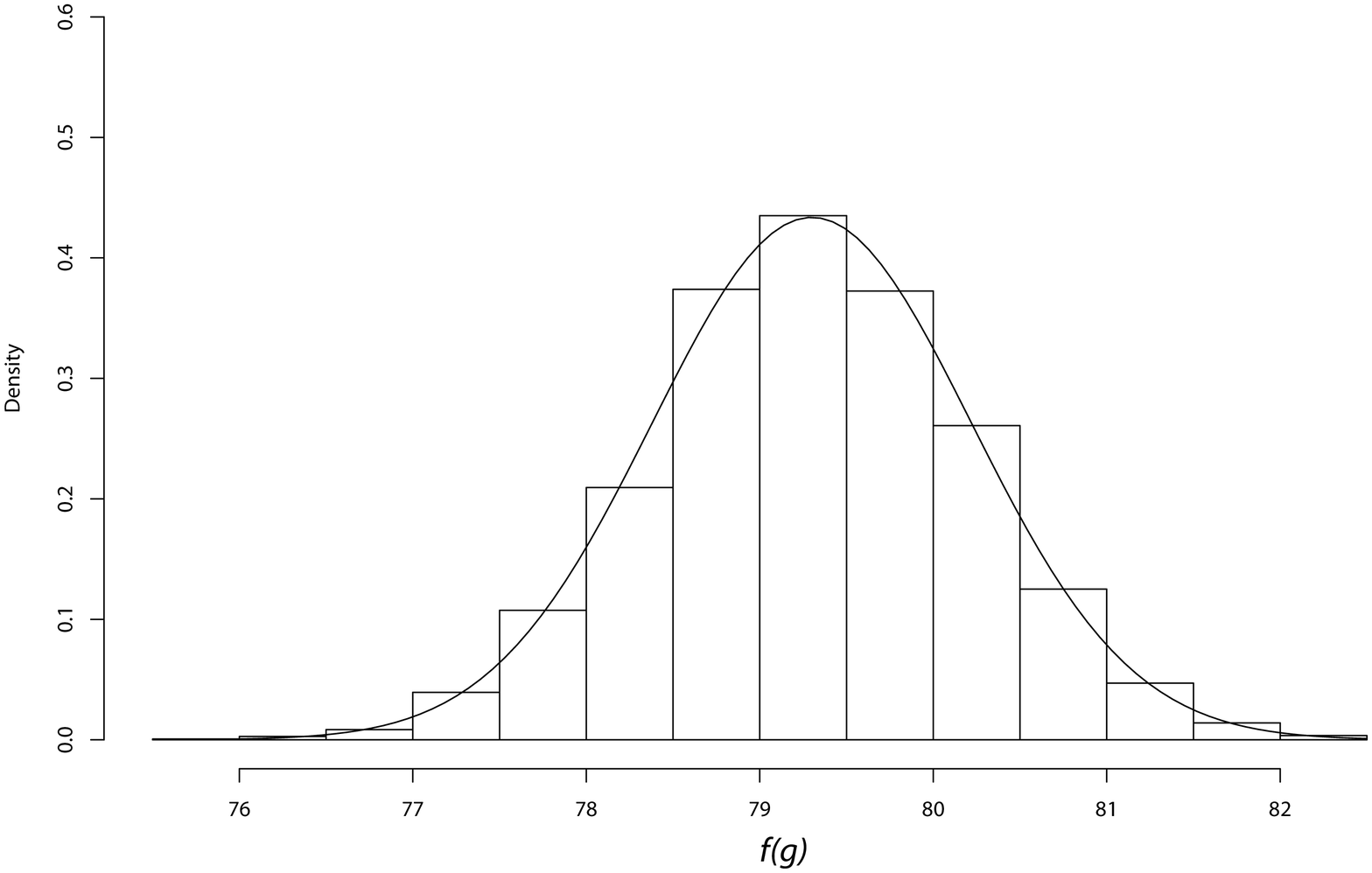}}~
    \label{gaussian}
  }
 \caption{
The value of the action as a function of the (logarithm of the) evolution time $\tau$ for $g=20$ and the distribution 
at sufficiently late times.  
(a) The evolution of the value of the free energy along the Monte Carlo time. Each point represents the value of the 
free energy on the accepted field configuration at the end of each sequence of $n_T\simeq 10$ 
steps (an HMC trajectory). 
%
%
After some time (in this case $\ln\tau \sim 6$) the state ``thermalizes", {\it i.e.} the value of 
the free energy on the generated field configurations follow a normal distribution. 
(b) The value of the free energy and the corresponding error (and in fact of any other observable) is found by 
fitting a Gaussian on sufficiently many values after thermalization; the histogram is constructed from about 500 data points.   
}
  \label{error_analysis}
\end{figure}

The universal scaling function for each value of $g$ is the result of an independent simulation. 
Following the RHMC algorithm \cite{Kennedy:1998cu, Clark:2003na, Clark:2006wq} reviewed in 
Appendix~\ref{RHMC}, field configurations are generated by starting with a random field configuration 
and evolving it along a fictitious time direction.  At the end of every 
$n_T$-step "time" sequence \footnote{The number of steps $n_T$ and the 
length of each step are tunable parameters chosen to decrease the thermalization time, see Figure~\ref{thermalization}.} 
the resulting field configuration is kept or rejected whether or not it can be interpreted as the result of an actual quantum 
mechanical evolution 
of the system and momenta are re-generated. 
It was shown in \cite{Kennedy:1998cu} that the 
resulting field configurations sample the complete phase space of the system.
After a sufficiently long evolution the values of the free energy (or of any other observable) follow a Gaussian distribution;
the free energy and its error are extracted as the mean value and the standard deviation of this distribution, 
respectively. 
Figure~\ref{thermalization} shows, for $g=20$, the evolution of the value of the action along the evolution towards 
thermal equilibrium and figure~\ref{gaussian} shows the corresponding Gaussian fit.\footnote{Here and for the 
other values of $g$  we carry out the fit using 
the Maximum-likelihood Fitting of Univariate Distributions in the statistical data analysis package R \cite{Simone:2011zz}.
See also \tt{http://stat.ethz.ch/R-manual/R-devel/library/MASS/html/fitdistr.html}}

\begin{figure}
  \centering
  \subfigure[]{%
    \includegraphics[scale=0.73]{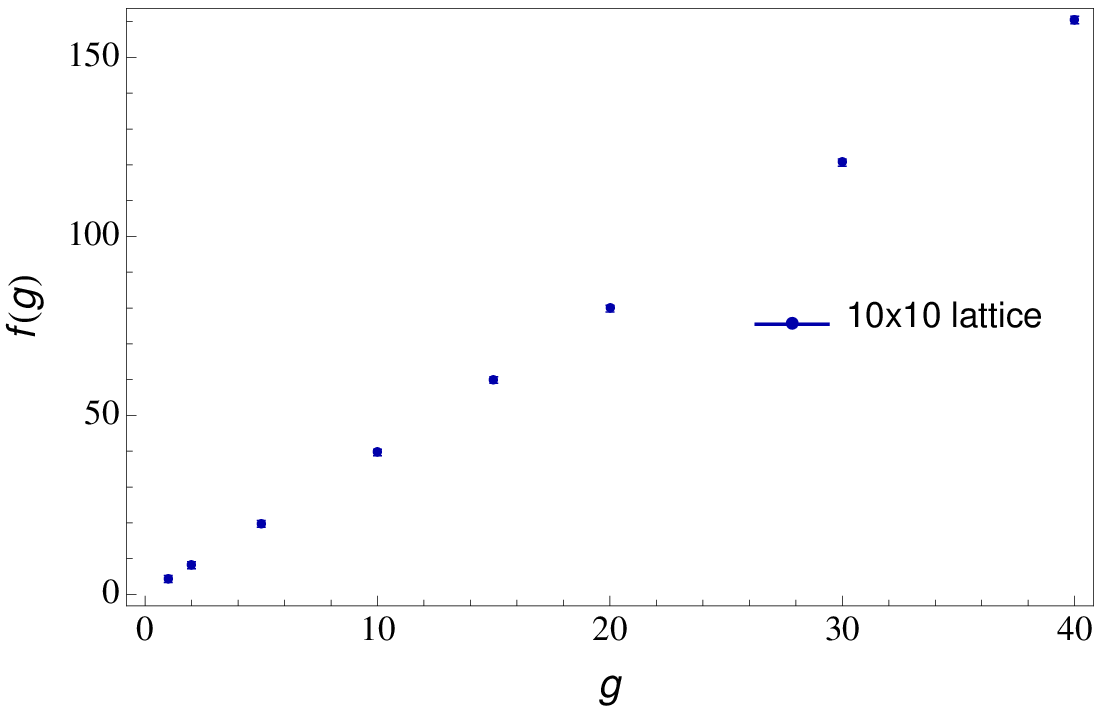}
    \label{10x10}
  }
  \subfigure[]{%
    \includegraphics[scale=0.73]{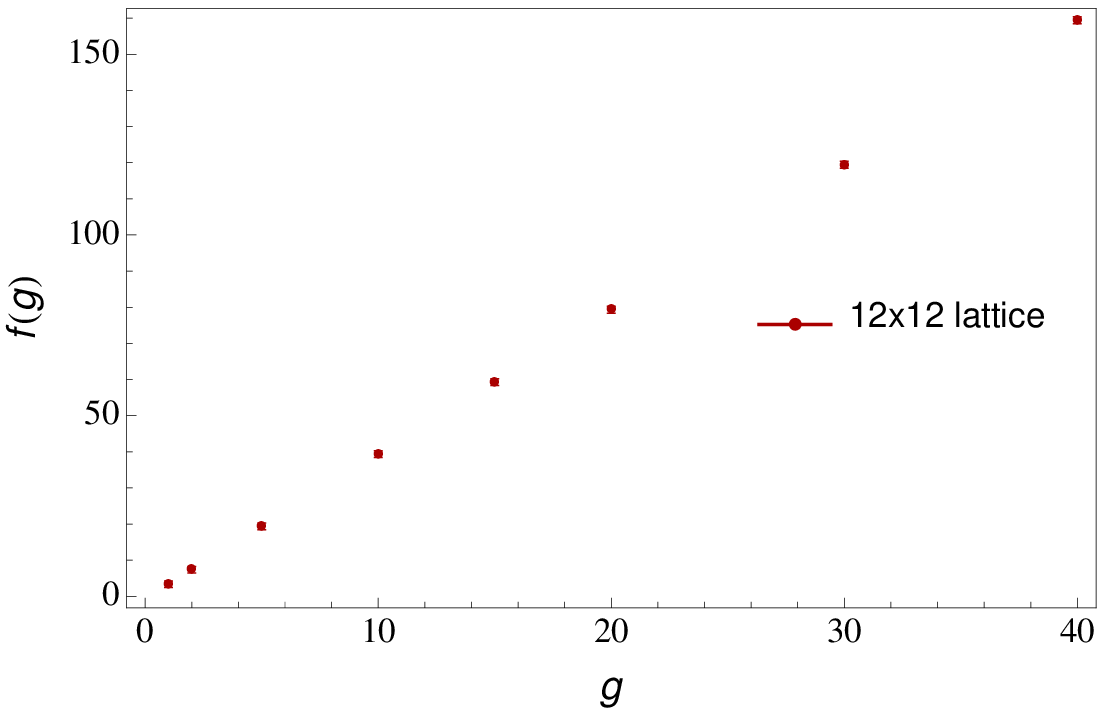}
    \label{12x12}
  }
  \caption{The universal scaling function for the values of $g$ in eq.~\eqref{gvalues}
  from  $10\times 10$ and $12\times 12$ lattice simulations.
  }
  \label{data_fits_10_12}
\end{figure}

The results of the simulations of the action \eqref{final_continuum_L}  on a $10\times 10$ and $12\times 12$
lattice are listed in the second and third two columns of  table~\ref{tab:data} and plotted in figures \ref{10x10}
and \ref{12x12}, respectively. On the scale of the plot the  two data sets are practically indistinguishable.
Inspecting the numerical values reveals small variations in the position of the central value as well as a
reduction in their absolute errors. This reduction is consistent with the expectation that the discretization errors
are smaller on finer lattices.
In the fourth column of table~\ref{tab:data} we include the values of the universal scaling function obtained
by solving the BES equation for the same values of $g$.\footnote{We thank D.~Volin for a numerical solution 
of the BES equation.}

\begin{table}[h]
\centering 
\begin{tabular}{| r | r | r | r |} 
\hline
$\vphantom{a^{\big|}}g$ & $10 \times10$ lattice &  $12\times12$ lattice  &  BES equation  \\ [0.5ex]
\hline 
1	&$4.328\pm 0.97$ 		& $3.335\pm .89$ 		&  3.3066         \\
2	&$8.127\pm 0.96$		& $7.385\pm .91$		&  7.3246		\\
5	&$19.694\pm 0.92$		& $19.373\pm .91$		& 19.3332		\\
10	&$39.652\pm 0.92$		& $39.380\pm .92$		& 39.3359		\\
15	&$59.878\pm 0.93$		& $59.306\pm .93$		& 59.3369		\\
20	&$79.804\pm 0.95$		& $79.313\pm .94$		& 79.3429		\\
30	&$120.628\pm 1.00$		& $119.408\pm .97$		&119.4052	\\
40	&$160.480\pm 1.05$ 	 & $159.422\pm .98$		&159.5485	\\
\hline 
\end{tabular}
\caption{The numerical values of the universal scaling function obtained from
 $10\times 10$ and $12\times 12$ lattices as well as the results of the BES equation. The latter
 are quoted with an uncertainly of one unit in the last digit. \label{tab:data}}
\end{table}

The sources of errors are well-understood, see Appendix~\ref{RHMC_errors} for a brief summary. The lower bound on the finite 
volume error estimated in \eqref{lower_bound0} becomes here
\be
\Delta f(g) \ge \frac{4}{V_2} =\frac{4}{\pi^2} \simeq 0.4 \ .
\label{lower_bound}
\ee
This accounts for about $50\%$ of the reported error in table \ref{tab:data}.  
%
At large values of the coupling constant the extra error is statistically-dominated due to a slow thermalization time
and low acceptance rate of the RHMC algorithm; this can presumably be justified by the fact that at large values of $g$
the partition function is dominated by a single classical field configuration rather than by a distribution of field 
configurations.
At small values of $g$ the error induced by the fermion determinant  provides the bulk of the extra error; this is 
a consequence of the fact that fermions become effectively light in this regime.
It is difficult to estimate the precise effects of the discretization and of the finite lattice spacing; by comparing the 
values in the second and third columns we see that the error decreases with the decrease of the lattice spacing, in 
agreement with expectations and implying that even finer lattices will yield higher-precision results.
We also ran these simulations for a smaller volume \footnote{Increasing 
the lattice volume at fixed lattice spacing is computationally very expensive.} and found a substantial increase of the 
error estimate.  This is consistent with an increase in the lower bound on the finite-volume uncertainty 
estimate~\eqref{lower_bound}.

\begin{figure}[tH!]
\centering{
\includegraphics[scale=0.85]{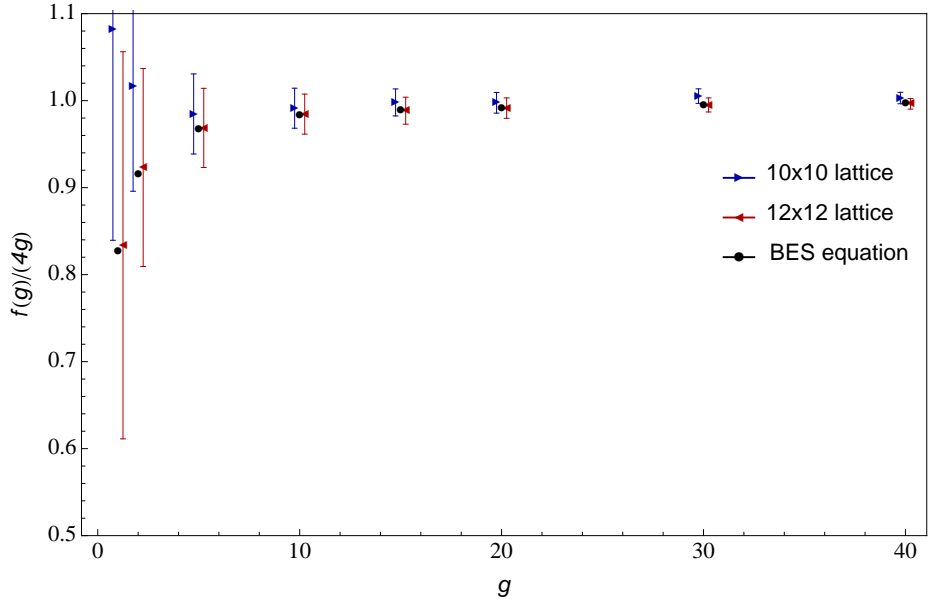}
}
\caption{Plot of the rescaled universal scaling function from the $10\times 10$ (blue right-triangles) and 
$12\times 12$ (red left-triangles) and its values
from the BES equation (black dots). The lattice values are artificially displaced by $\delta g = \pm1/4$ for easy comparison.
Clearly, the central values of the $12\times12$ lattice is a very good approximation of the integrability results.
\label{comparison}}\nonumber
\end{figure}

In figure~\ref{comparison} we have plotted the three data sets rescaled by
a factor of $(4g)$ such that at large values of $g$ the graph asymptotes to one; to facilitate the comparison we
also artificially shifted the plot along the horizontal axis by $\delta g = -1/4$ for the $10\times 10$ lattice data and by
$\delta g = +1/4$ for the $12\times 12$ lattice data while keeping fixed the values obtained from the BES equation.
We notice a very good agreement between the BES (black dots) and the central values of the $12\times 12$ lattice 
results (red triangles) while still being in fair agreement within the error bars with the $10\times 10$ lattice results (blue 
triangles). 
%
%

\begin{table}[h]
\centering 
\begin{tabular}{|c|r| r|r|} 
\hline 
Origin $\vphantom{a^{\big|}}$ & $a_0$ & $a_1$ & $a_2$ \\ [0.5ex]
\hline 
$\vphantom{^{A^A}}$$10\times 10$ lattice & $4.031\pm 0.007$ &$-0.667 \pm 0.188$ & $-$ \\ 
$12\times 12$ lattice & $4.001\pm 0.002$ & $-0.662  \pm 0.041$ & $-0.015 \pm 0.066$ \\
perturbation theory& $4.000\pm 0.000$  & $-0.662\pm 0.001$         &  $-0.023\pm 0.001$\\ 
\hline 
\end{tabular}
\caption{Coefficients of the fit of the lattice data with the expected form of the worldsheet perturbative expansion.}
\label{tab:fit_coefs}
\end{table}

While the reported absolute errors for universal scaling functions are relatively large, we can try to make
contact with worldsheet perturbation theory by fitting the lattice data in table~\ref{tab:data}
onto the known form of the worldsheet perturbation theory,
\be
f(g)=a_0\, g + a_1 +\frac{a_2}{g}+\dots \ .
\ee
The resulting coefficients and their errors are listed in table~\ref{tab:fit_coefs}. Not surprising, the quality
of the fit degrades as one attempts to extract higher-order coefficients ({\it e.g.} an estimate for the two-loop
coefficient cannot be extracted reliably from the $10\times 10$ lattice). This is consistent with the observation 
that the first two terms in worldsheet perturbation theory provide a good approximation to the solution of 
the BES equation for $g>1$.
The fit may be slightly improved by assuming an expansion in $\sqrt{1+16g^2}$ rather than in $g$; this accounts for the
expected $g_*=1/4$ radius of convergence of sYM perturbation theory. In particular, the central value of $a_2$ becomes
very close to the results of perturbation theory at the expense of a slightly poorer fit for $a_1$.

\begin{table}[h]
\centering 
\begin{tabular}{|c|r| r|r|} 
\hline 
Origin$~\big{\backslash}~~ g$ $\vphantom{a^{\big|}}$ & ${10^{-1}}$ & $({4\sqrt{\pi}})^{-1}$ & ${4}^{-1}$ \\ [0.5ex]
\hline 
$12\times 12$ lattice & $0.014\pm 0.67$ & $0.095\pm 0.69$ & $0.581\pm 0.82$ \\
BES equation            & $0.077\pm 10^{-3}$  & $0.150\pm 10^{-3}$         &  $0.427\pm 10^{-3}$\\[1ex] 
\hline 
\end{tabular}
\caption{Values of the universal scaling function inside the radius of convergence of $\NeqFour$ sYM theory.}
\label{tab:small_g}
\end{table}

We have also evaluated the universal scaling function for values of $g$ at or below the expected radius of convergence
of $\NeqFour$ sYM perturbation theory, $g_* = 1/4$. The results are included in table~\ref{tab:small_g}. Since expected
values for $f(g)$ are close to or below the expected lower bound of the error of the simulation on a lattice of volume
$V_2=\pi^2$, \eqref{lower_bound}, the results cannot be statistically significant; we nevertheless note that, close to
$g=g_*$, where the value of the cusp anomaly is larger than the error's lower bound, the central value is relatively close
to the result of the BES equation. It should be possible -- albeit nontrivial -- to reduce the error bar on such data points.
We also expect that further reducing the lattice spacing will lead to the central value moving closer to the BES prediction.

\section{Summary and further comments  \label{conclusions_and_more}}

In this paper we computed the energy of the folded string in \adss~ (and thus the universal scaling function) 
at finite values of the 't~Hooft coupling using the Green-Schwarz string in \adss~ by discretizing the worldsheet 
theory in the relevant background and computing numerically its partition function. 
Our results reproduce the predictions of the Asymptotic Bethe Ansatz within the accuracy of our simulation 
and thus strongly support the expectation 
that the Asymptotic Bethe Ansatz yields the long string spectrum, or the spectrum of long operators in the dual 
$\NeqFour$ sYM theory for all values of the coupling. 
We have attempted to find the universal scaling function inside the radius of convergence of $\NeqFour$ sYM theory;
while our simulation is not sufficiently precise for this purpose, there is no conceptual obstacle.\footnote{These methods 
are however inappropriate for studying the regime $\lambda<0$, whose existence is the reason for a finite radius of 
convergence for the planar theory.} 

Our calculation can in principle be extended to other long string states as well as to short string states that have 
a description as the small charge continuation of long string states. The computational complexity depends on the
details of the state and is strongly correlated with the presence or absence of massless fermions. 
We have also discussed strategies for finding the spectrum of general string states that are not in this class
from the two-point functions of certain worldsheet (vertex) operators. 
By computing the two-point function of worldsheet fluctuations in long string background it should be possible to 
find information on their spectra at finite values of the coupling.

The sources of errors in such calculations are well-understood; it turns out however that the single dominant source -- 
related to the need to use a lattice of finite extent in both space-like and (Euclidean) time-like directions -- is the most 
difficult to overcome. A volume increase at fixed lattice spacing requires an increase in the number of lattice sites which 
increases the duration of the calculation. A second -- but not less important -- consequence of the increase in the number 
of lattice sites is the rapid increase in the size of the fermion matrix ($16^2$ times faster than the increase in the number 
of lattice sites, due to the number of independent fermions); this in turn leads to  either an increase in the time needed 
to find a solution to the systems \eqref{systems} (necessary for the generation of field configurations) or a decrease in 
the accuracy of the solution. \footnote{Perhaps more dramatically, as the size of the matrix increases so does the memory needed 
to store it.  One of the most expensive operations in the GPU calculations is data transfer;
if the  matrix is sufficiently large such that it cannot be fully stored on the GPU, most of the advantage of 
the GPU speedup is lost. }
It therefore appears that perhaps the most efficient improvement of our simulation it to employ a more efficient
treatment of the fermion matrix. At the analytical level one may consider factorizing it into simpler ({\it e.g.} upper-triangular 
and lower-triangular) factors and exponentiating each factor separately. At the level of the numerical calculation 
one could attempt to use a Fourier-accelerated (R)HMC algorithm \cite{Catterall:2001jg}, which may reduce the thermalization 
time.
One can also attempt to use a more efficient algorithm for solving the systems \eqref{systems} ({\it e.g.} one that accepts a 
preconditioner -- the numerical analog of the analytic factorization). Moreover, to alleviate the issues related to the size 
of the fermion matrix one could attempt to use a parallel GPU solver. 
A further possibility is to try to construct lattice actions that are much less sensitive to the lattice discretization than the one 
we used, perhaps along the lines of results in bosonic two-dimensional sigma models \cite{Balog:2012db}. 
If such actions exist  for the Green-Schwarz string they would allow use of larger lattice spacing while maintaining accuracy and 
thus would allow larger volume latices.

The physically very interesting case of short string states provides a natural partial solution to the difficult 
consequences of a larger lattice 
volume. Indeed, the worldsheet of such strings is a cylinder and thus increasing the volume requires increasing only 
one of the two dimensions of the lattice. Consequently, the increase in the number of lattice sites occurs at a much 
slower rate allowing in principle for smaller finite-volume errors while simultaneously keeping the discretization and 
statistical errors under control. It would be very interesting analyze states on the first excited string level using the 
methods proposed in this paper.

\bigskip

\section*{Acknowledgments }

We would like to thank Joe Polchinski and Arkady Tseytlin for useful discussions. 
We would also like to thank Dima Volin for providing a numerical solution for the BES equation.
This work is supported in part by the US Department of Energy under contract DE-SC0008745.

\bigskip


\begin{appendices}

\section{Simulating the lattice: Algorithms \label{RHMC}  }

\subsection{The Rational Hybrid Monte Carlo  algorithm}

In this appendix we review the structure of the Rational Hybrid Monte Carlo (RHMC) algorithm~\cite{Kennedy:1998cu,
Clark:2003na, Clark:2006wq} which we used to simulate the Green-Schwarz string. It differs from the standard
Hybrid Monte Carlo (HMC) algorithm~\cite{Duane:1987}  in the treatments of the fermion contribution for which it uses
a rational approximation for the fractional power of the quadratic fermion matrix:
\be
(M^{\dagger}M)^{-\frac{1}{4}} = \alpha_0+\sum_{i=1}^{P}\frac{\alpha_i}
{M^{\dagger}M+\beta_i}
\label{rational_approx}
\ee
with real $\beta_i$. This approximation is obtained though the Remez algorithm \cite{Remez, Clark:2006wq} implemented 
{\it e.g.} in {\tt{alg$\_$remez}} which is part of the library {\tt{RHMC-on-GPU}}.

\subsection{Monte Carlo methods}

A method to evaluate high-dimensional integrals such as the path integrals necessary to evaluate expectation values
of operators in quantum field theories with fields $\phi$ and Euclidean action $S_E[\phi]$
\be
\langle {\cal O}(\phi) \rangle =\frac{1}{Z}\int D\phi\; {\cal O}(\phi)\,e^{-S_E[\phi]}
\ee
is to randomly generate a sequence of field configurations with probability $P(\phi) = 1/Z\exp(-S_E[\phi])$ and then
construct the ``time" average
\be
{\overline {\cal O}} = \frac{1}{T}\sum_{i=1}^T {\cal O}(\phi_i) \ .
\label{average}
\ee
In the limit $T\rightarrow\infty$ the statistical expectation value $\langle {\cal O}(\phi) \rangle $ and the time-average
${\overline {\cal O}} $ are equal up to corrections ${\cal O}(T^{-1/2})$.
A similar method is used to construct the partition function except that the normalization factor of the probability,
$P(\phi) = {\cal N} \exp(-S_E[\phi])$, is chosen on physical grounds.
A useful algorithm uses a Markov process which generates a new field configuration $\phi'$ from the old configuration $\phi$
with probability $P_M(\phi\rightarrow\phi')$ which samples the entire configuration space and satisfies
\be
P_S(\phi)P_M(\phi\rightarrow\phi') = P_S(\phi') P_M(\phi'\rightarrow\phi) \ .
\ee
These conditions guarantee convergence to a unique distribution $P_S$.

It is convenient to split the generation of the new field configuration in two steps: (1) one generates a new field configuration
from the old one by some method and with some probability $P_C$ and (2) one chooses between the  newly generated
configuration and the old one (to be called ``new configuration" if chosen) with some probability $P_A$. Any information
about  the initial state will be lost after a sufficiently large number of steps.
In this construction it is important that correlations between successive configurations be minimal; moreover, for the
process to sample sufficiently quickly large parts of configuration space, it is useful to have a relatively large acceptance
probability $P_A$, of the order of $50-90\%$.

\subsection{The HMC algorithm}

An elegant method which realizes these ideas is the Hybrid Monte Carlo (HMC) algorithm proposed in ref.~\cite{Duane:1987};
a deterministic molecular dynamics evolution is used to generate new field configurations and a stochastic Metropolis
acceptance test is used to select the configurations which are retained; see {\it e.g.} \cite{HMCbookchapter} for a thorough discussion
of this algorithm and its variations. 

Denoting by the index $\phi$ and $\zeta$ the bosonic fields and the pseudo-fermions (cf. eq.~\eqref{fermiondet}),
ref.~\cite{Duane:1987} postulates the Hamiltonian
\be
H_\tau= \frac{1}{2}\pi_\phi^2 + {\bar\pi}_{\zeta} \pi_{\zeta} + S_{\phi}+ S_\zeta \ ,
\label{evolution_H}
\ee
which describes the evolution along some fictitious direction $\tau$. All fields are assumed to depend on this fictitious coordinate.
The new fields $\pi_\phi$ and $\pi_\zeta$ are interpreted as momenta conjugate to bosons and pseudo-fermions. The partition
function with this Hamiltonian is the partition function with the action $S_b+S_\zeta$  up to a normalization factor from
integrating out $\pi_\phi$ and $\pi_\zeta$. This factor may be found by computing the partition function with $S_\phi+S_\zeta=0$

\begin{figure}[htb]
   \centering
   \includegraphics[scale=0.6]{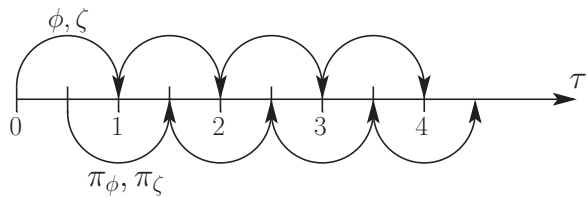}
   \caption{Sketch of the Leapfrog method: where values of fields  and their conjugate momenta ``leap" over each other.}
   \label{fig:leapfrog}
\end{figure}

Given some initial field configuration $(\phi, \zeta)$ and some randomly generated momenta $\pi_\phi$ and $\pi_\zeta$
a new field configuration is constructed deterministically by solving Hamilton's equations of motion for $H_\tau$:
\be
\partial_\tau {\pi}_\phi &=&-\frac{\partial H_\tau}{\partial \phi}\equiv F_\phi \quad\qquad\qquad    \quad
\partial_\tau{\pi}_\zeta =-\frac{\partial H_\tau}{\partial \zeta} \equiv  F_\zeta
\\
\partial_\tau {\phi}&=&\frac{\partial H}{\partial \pi_\phi} = \pi_\phi \quad\qquad\quad ~
\partial_\tau {\zeta}=\frac{\partial H}{\partial \pi_\zeta} = {\bar\pi}_\zeta \ ,
\ee
where $F_\phi$ and $F_\zeta$ are bosonic and fermionic ``forces".
These equations are integrated with an energy-conserving symplectic ({\it i.e.} phase space-area preserving)
integrator that uses the standard leapfrog scheme:
we update the values of fields and their conjugate momenta at staggered time steps, such that one is leaping over
the other, see figure~\ref{fig:leapfrog}. For the bosons $\phi$ the evolution over $\delta \tau$ is given by
\be
\pi_{\phi,\tau+\delta \tau/2}&=& \pi_{\phi,\tau} +  F_{\phi, \text{initial}}  \frac{\delta \tau}{2}
\quad,\quad
\phi_{\tau+\delta \tau}= \phi_\tau + \pi_{\tau+\delta \tau/2} \delta \tau \\
\pi_{\phi;\tau+\delta \tau}&=& \pi_{\phi;\tau+\delta \tau /2} +  F_{\phi;\text{updated}}  \frac{\delta \tau }{2} \ .
\label{leapfrog}
\ee
The evolution of $\zeta$ and
$\pi_\zeta$  is similar. This integration method introduces only an ${\cal O}(\delta\tau^2)$ error.\footnote{ 
To see this  \cite{Sexton:1992nu} one compares the infinitesimal evolution with $H_\tau=T(\pi_\phi, \pi_\zeta)+S(\phi, \zeta)$ 
with the step-wise evolution operator \eqref{leapfrog}, 
$$
e^{-(S+T)\delta\tau}\quad \text{vs.}\quad e^{-\frac{1}{2}\delta\tau S}e^{-\delta\tau T}e^{-\frac{1}{2}\delta\tau S} \ .
$$
Using  the Baker-Campbell-Hausdorff relation
\be
e^X e^Y &=& e^{C(X,Y)}\qquad
C(X,Y)=C_1(X,Y)+C_2(X,Y)+C_3(X,Y)+\dots
\cr
C_1&=& X+Y \quad C_2= \frac{1}{2} [X,Y] \quad C_3= \frac{1}{12}([[X,Y],Y]-[[X,Y],X])
\nonumber
\ee
and the fact that the action of $S$ and $T$ is given by the Poisson brackets 
of momenta and fields, the two evolution operators can be easily related:
$$
e^{-\frac{1}{2}\delta \tau {S}}e^{-\delta \tau {T}}e^{-\frac{1}{2}\delta \tau\, {S}} \simeq \exp\{-\delta \tau ({S} + {T}) 
- \delta \tau^3 ([[{S},{T}],{S}]+[[{S},{T}],{T}]) +O(\delta \tau^5)\} \ .
$$
The antisymmetry of the Poisson bracket leads to the cancellation of the contribution of the commutator $C_2$ in the Baker-Campbell-Hausdorff formula. Thus, the
two evolution operators differ by terms suppressed by a factor of $\delta\tau^2$.
}

To eliminate this artificial error the field configuration obtained after some number $n_T$ of time steps -- known as 
an HMC trajectory -- 
undergoes a  Metropolis acceptance step \cite{Duane:1987}: a number $n_M\in[0, 1]$ is randomly generated and the final 
configuration is accepted if  $n_M < e^{- \delta H_\tau}$ where $\delta H_\tau$ is the change in the value of the 
Hamiltonian \eqref{evolution_H} between the initial and final field configuration of that trajectory.
If  $e^{- \delta H_\tau}<n_M$  then the initial field configuration is accepted as the new field configuration contributing to
\eqref{average}. At the beginning of each HMC trajectory -- {\it i.e.} after each Metropolis test -- momenta are refreshed 
based on a Gaussian distribution (centered at zero and with unit standard deviation), which will keep the simulation ergodic.
Since in the limit $\delta\tau\rightarrow 0$ the integrator preserves the energy, the only reason $e^{- \delta H}$ might not equal
unity is the presence of some error due to the finite time-step $\delta\tau$. The inclusion of the Metropolis acceptance step
renders the HMC algorithm exact, with results independent of step size  \cite{Duane:1987} (reversibility of the dynamics is
important for proving this).

\subsection{The fermion contribution to bosonic RHMC forces}

Due to the rational approximation \eqref{rational_approx} of $(M_\text{string}^{\dagger}M_\text{string})^{-\frac{1}{4}}$, 
the fermionic contribution to the bosonic forces takes a relatively simple form, which suggests an efficient way to
evaluate it. Denoting as before a generic boson by $\phi$ and the pseudo-fermions by $\zeta$, the pseudo-fermion
contribution to the bosonic forces $F_\phi^{\zeta} $ is:
\begin{align}
F_\phi^{\zeta} = -\frac{\partial S_\zeta}{\partial \phi}
&=  \sum_{i=1}^P \alpha_i (\frac{1}{ M^{\dagger}M+\beta_i}\zeta)^{\dagger} \frac{\partial}{\partial \phi}(M^{\dagger}M) (\frac{1}{ M^{\dagger}M+\beta_i} \zeta) \nonumber \\
&=  \sum_{i=1}^P \Big[\alpha_i \big(M\frac{1}{ M^{\dagger}M+\beta_i}\zeta\big)^{\dagger} \frac{\partial M }{\partial \phi}
\big(\frac{1}{ M^{\dagger}M+\beta_i} \zeta\big) \nonumber \\
&\qquad\qquad\qquad\qquad
+  \alpha_i \big(\frac{1}{ M^{\dagger}M+\beta_i}\zeta\big)^{\dagger} \frac{\partial M^{\dagger} }{\partial \phi}
\big(M\frac{1}{ M^{\dagger}M+\beta_i}\zeta\big)\Big]\nonumber \\
&=  \sum_{i=1}^P \alpha_i \left((M s_i)^{\dagger}\frac{\partial M }{\partial \phi} s_i + s_i^{\dagger}\frac{\partial M^{\dagger} }{\partial \phi} M s_i \right) \ ,
\end{align}
where $s_i =  {1}/{(M^{\dagger}M+\beta_i)} \zeta$ are interpreted as the solutions to the matrix equation
\be
(M^\dagger M +\beta_i)s_i = \zeta \quad,\qquad i=1,\dots,P \ .
\label{systems}
\ee
In the discretized theory $\zeta$ stands for the vector of values of the field $\zeta$ at all lattice sites, $M^\dagger M$ is a
matrix of size $(\text{nr. of fermion components}\times\text{nr. of lattice sites})^2$. For example, for the
cusp anomaly calculation in section~\ref{explicitcalculation} on the $12\times 12$ lattice, $M^\dagger M$ has size
$(16*12^2)\times(16*12^2)$ and $\zeta$ stands for a $(16*12^2)$-component vector comprising all the components ($16$)
pseudo-fermions at all lattice sites ($12^2$). 
%

The systems \eqref{systems} are solved using a multi-mass conjugate gradient solver -- in particular {\tt{gc$\_$m}} routine 
which is part of the CUSP library \cite{Cusp} -- the  which allows for the solution of all $P$ systems in a single solve. 
As long as all the shifts $\beta_i$ are positive one can solve
for the entire family of solutions at the cost of solving for a single unshifted system: one solves for the shifted system 
with slowest convergence\footnote{$\beta_i$=0 is the slowest-converging system.} and then the other shifted
solutions can be found by an additional multiplication step~\cite{Jegerlehner:1996pm}. The standard  conjugate gradient
solver is described in \cite{Hestenes}.

With the same notation the fermionic forces are:
\be
F_\zeta= - \frac{\partial S_\zeta}{\partial \zeta} &=& -\alpha_0 \frac{\partial}{ \partial \zeta} (\zeta^{\dagger} \zeta) - \sum_{i=1}^P
\frac{\partial}{\partial \zeta}(\zeta^\dagger \frac{\alpha_i}{ M^{\dagger}M+\beta_i}\zeta)
=-\alpha_0 \zeta^{\dagger} - \sum_{i=1}^P \alpha_i s_i^{\dagger} \ 
\ee
and are determined by the solution to the same system \eqref{systems}.
Even though the fermionic matrix, $M^{\dagger} M$, is not guaranteed to be symmetric, its hermiticity guarantees 
that the multi-mass conjugate-gradient method will yield a solution to \eqref{systems} \cite{Ten:2010as, Hestenes}.

\subsection{A summary of sources of errors \label{RHMC_errors}}

The RHMC algorithm is considered to be an exact algorithm to machine precision \cite{Duane:1987}. Nevertheless, 
errors are introduced through the approximations made in the construction of the simulation; they have both a statistical 
and a systematic origin. We summarize here the ones relevant for the Green-Schwarz string in \adss~ space:

\begin{itemize}

\item Statistical error -- arises because the path integral (with or without additional operator insertions) is approximated in terms
of a finite number of field configurations. In our calculations we employed ${\cal O}(500)$ independent configurations.
Statistical errors may be reduced by increasing the number of field configurations.

\item Discretization errors  -- arise because of the finite lattice spacing and in the extrapolation to the continuum limit.
In two dimensions the approach to the continuum limit is accompanied by a quadratic increase in the number of lattice 
sites which in turn leads to an increase in the computational cost at a higher rate. These errors may be reduced by 
employing higher-point approximations for field derivatives.

\item Other approximation errors -- arise because of the Remez algorithm used to construct a rational approximation of 
the inverse fractional power as well as because of the numerical errors of the multi-mass conjugate-gradient solver. 
The former may be reduced by using more terms in eq.~\ref{rational_approx}. The latter may potentially be reduced 
by improving the treatment of the fermion matrix, such as using a preconditioner and other algorithms for solving the 
linear systems \eqref{systems}.

\item Finite volume errors -- all simulations are carried out on lattices of finite extent. 
Since the Compton wave length of a massive particle is 
proportional to its inverse mass,  finite volume effects are larger for particles of smaller masses. 
In the context of our calculations 
we have estimated these errors in sec.~\ref{action_and_stuff}.

\end{itemize}

\section{9-point stencil \label{9ptstencil}  }

We want to construct a discrete approximation to the derivative of some function $f(x)$ such that the error is of the order
${\cal O}(a^8)$ where $a$ is the lattice spacing. This requires a nine-point stencil using
$f(x\pm a),f(x\pm 2a),f(x\pm 3a),f(x\pm 4a)$.  We expand then in Taylor series around $a=0$ to ${\cal O}(a^8)$ and
then solve the resulting system for $\partial_x f(x)$:
\begin{eqnarray}
f(x\pm a)&=& f(x) \pm a f^{(1)}(x) + \frac{1}{2}a^2 f^{(2)}(x) \pm \frac{1}{3!} a^3 f^{(3)}(x) + \frac{1}{4!} a^4 f^{(4)}(x) \\
&&\pm \frac{1}{5!} a^5 f^{(5)}(x) + \frac{1}{6!} a^6 f^{(6)}(x) \pm \frac{1}{7!} a^7 f^{(7)}(x) + {\cal O}(a^8) \\
f(x\pm 2a)&=& f(x) \pm 2 a f^{(1)}(x) + 2 a^2 f^{(2)}(x) \pm \frac{4}{3} a^3 f^{(3)}(x) + \frac{2}{3} a^4 f^{(4)}(x) \\
&&\pm \frac{2^5}{5!} a^5 f^{(5)}(x) + \frac{2^6}{6!} a^6 f^{(6)}(x) \pm \frac{2^7}{7!} a^7 f^{(7)}(x) + {\cal O}(a^8) \\
f(x\pm 3a)&=& f(x) \pm 3 a f^{(1)}(x) + \frac{9}{2} a^2 f^{(2)}(x) \pm \frac{9}{2} a^3 f^{(3)}(x) + \frac{27}{8} a^4 f^{(4)}(x) \\
&&\pm \frac{3^5}{5!} a^5 f^{(5)}(x) + \frac{3^6}{6!} a^6 f^{(6)}(x) \pm \frac{3^7}{7!} a^7 f^{(7)}(x) + {\cal O}(a^8) \\
f(x\pm 4a)&=& f(x) \pm 4 a f^{(1)}(x) + 8 a^2 f^{(2)}(x) \pm \frac{32}{3} a^3 f^{(3)}(x) + \frac{32}{3} a^4 f^{(4)}(x) \\
&&\pm \frac{4^5}{5!} a^5 f^{(5)}(x) + \frac{4^6}{6!} a^6 f^{(6)}(x) \pm \frac{4^7}{7!} a^7 f^{(7)}(x) + {\cal O}(a^8)
\end{eqnarray}
Requiring that
\begin{eqnarray}
\alpha \,a\, \partial_x f(x) + {\cal O}(a^9) &=&
e (f(x + a) - f(x - a)) - b (f(x + 2a) - f(x - 2a))
\\
&&\qquad
 - c (f(x + 3a) - f(x - 3a))  - d (f(x + 4a) - f(x - 4a))
 \nonumber
\end{eqnarray}
constrains the coefficients on the right-hand side to be a solution of the system
\be
e - 8 b - 27 c - 64 d &=&0
\cr
e - 32 b - 243 c - 1024 d &=&0
\\
e - 128 b - 2187 c - 16384 d &=&0
\nonumber
\ee
The solution $e=224, b=56, c=-\frac{32}{3}, d=1$ leads to $\alpha = 280$
and to the nine-point stencil quoted in eq.~\eqref{discrete_derivative_9pt}.

\subsection{$\rho$ matrices}

We include here the explicit form of the matrices $\rho^M$ entering the AdS light-cone gauge action.
They are off-diagonal blocks of the six-dimensional Dirac matrices in chiral representation:
\be
&&\rho_{ij}^M =- \rho_{ji}^M\,, \ \ \ \ \
   (\rho^M)^{il}\rho_{lj}^N + (\rho^N)^{il}\rho_{lj}^M
 =2\delta^{MN}\delta_j^i\,, \ \ \ \ \
  (\rho^M)^{ij}\equiv  -
 (\rho_{ij}^{M})^* \,
 \\
  &&
  \rho^{MN}{}^i{}_{ j} =\frac{1}{2}[  (\rho^M)^{il}\rho_{lj}^N
- (\rho^N)^{il}\rho_{lj}^M ]    \ .
\ee
One  can
choose  the following explicit  representation for the
$\rho^M_{ij} $ matrices
\be
\rho^1_{ij}&=&\left(\begin{matrix}0&1&0&0\\-1&0&0&0\\0&0&0&1\\0&0&-1&0
\end{matrix}\right)\,,\qquad
\rho^2_{ij}=\left(\begin{matrix}0&\mathrm{i}&0&0\\-\mathrm{i}&0&0&0\\0&0&0&-\mathrm{i}\\0&0&\mathrm{i}&0
\end{matrix}\right)\,,\qquad
\rho^3_{ij}=\left(\begin{matrix}0&0&0&1\\0&0&1&0\\0&-1&0&0\\-1&0&0&0
\end{matrix}\right)\,,\nonumber \\
\rho^4_{ij}&=&\left(\begin{matrix}0&0&0&-\mathrm{i}\\0&0&\mathrm{i}&0\\0&-\mathrm{i}&0&0\\ \mathrm{i}&0&0&0
\end{matrix}\right)\,,\qquad
\rho^5_{ij}=\left(\begin{matrix}0&0&\mathrm{i}&0\\0&0&0&\mathrm{i}\\-\mathrm{i}&0&0&0\\0&-\mathrm{i}&0&0
\end{matrix}\right)\,,\qquad
\rho^6_{ij}=\left(\begin{matrix}0&0&1&0\\0&0&0&-1\\-1&0&0&0\\0&1&0&0
\end{matrix}\right)\,,\nonumber
\ee

\section{Fermions \label{moreonfermions}  }

After the introduction of the auxiliary fields $\phi$ and $\phi_M$ the fermion Lagrangian is quadratic:
\be
L_{\text{fermions}} = \psi^T M \psi \ .
\ee
For the AdS light-cone gauge action  $\psi\equiv({\tilde\theta}^i, {\tilde \theta}_i, {\tilde\eta}^i, {\tilde\eta}_i)$ and
in the background of the null cusp solution $M$ is given by:
\be
\label{fermion-K}
&&M_\text{cusp}
\\
&=&\!\!
\begin{pmatrix}
0 &  {\rm i} \partial_t {\bf 1}_4 & -{{\rm i}} (\partial_s+\ha ) \rho^M \frac{{\tilde z}^M}{\tilde{z}^3} & 0 \cr
{\rm i} \partial_t {\bf 1}_4 & 0 & 0 & -{\rm i} (\partial_s+\ha ) \rho^{\dagger}_M \frac{{\tilde z}^M}{\tilde{z}^3}
\cr
{\rm i} \rho^M \frac{{\tilde z}^M}{\tilde{z}^3} (\partial_s-\ha )
& 0 & - \rho^M \frac{{\tilde z}^M}{\tilde{z}^4} (\partial_s {\tilde x} - \frac{{\tilde x}}{2} -{\tilde x} \partial_s ) & {\rm i} \partial_t {\bf 1}_4 + A^\dagger  \cr
0 & {\rm i} \rho^\dagger_M \frac{{\tilde z}^M}{\tilde{z}^3} (\partial_s-\ha )
& {\rm i} \partial_t {\bf 1}_4 + A & -\rho^\dagger_M \frac{{\tilde z}^M}{\tilde{z}^4} (\partial_s {{\tilde x}^*} - \frac{{{\tilde x}}^*}{2} -{{\tilde x}^*} \partial_s )
\end{pmatrix}
\nonumber
\ee
with $A$ given by
\be
A &=& \frac{1}{\tilde{z}^4} {\tilde\phi}^M \rho^{MN} {\tilde z}^N + \frac{1}{\tilde{z}^2} {\tilde\phi}
+\frac{{\rm i}}{\tilde{z}^2} {\tilde z}^N \rho^{MN}(\partial_t {\tilde z}^M) \ .
\ee

\end{appendices}


\bibliography{lattice_draft}
\bibliographystyle{nb}

%
%

\end{document}